\documentclass{article}
\usepackage{graphicx}
\usepackage{mathtools}
\usepackage{amsmath}
\usepackage{amssymb}
\usepackage{dsfont}
\usepackage{bbold}
\usepackage[a4paper, left=2cm, right=2cm, top=2cm, bottom=2cm]{geometry}
\usepackage{caption}
\usepackage{subcaption}
\usepackage{float}
\usepackage[normalem]{ulem}
\usepackage{microtype}
\usepackage{xargs}
\usepackage{xcolor}
\usepackage[colorinlistoftodos, prependcaption]{todonotes}
\newcommandx{\tdbd}[2][1=]{\todo[linecolor=red,backgroundcolor=red!25,bordercolor=red,inline,#1]{#2}}

\newcommandx{\tdls}[2][1=]{\todo[linecolor=rradiused,backgroundcolor=orange!25,bordercolor=orange,inline,#1]{#2}}

\usepackage[backend=biber, sorting=none, citestyle=numeric-comp, bibstyle=ieee, url=false, doi=false, isbn=false, eprint=false]{biblatex}
\AtEveryBibitem{%
  \clearfield{volume} 
  \clearfield{number} 
  \clearfield{pages} 
  \clearfield{note} 
  \clearfield{address} 
  \clearfield{month}
  \clearfield{day}
  \clearfield{doi}
}
\title{Localized and delocalized modes on random geometric graphs in 1D}

\author{Luca Schaefer and Barbara Drossel}
\date{Institute for Condensed Matter Physics, Technical University of Darmstadt,
 Hochschulstraße 6, 64289 Darmstadt, Germany}

\begin{document}
\maketitle
\abstract{We perform an extensive investigation of the localization properties of the eigenmodes of the Laplace and adjacency matrix for one-dimensional random geometric graphs. We evaluate the density of states, the probability distribution of the participation ratio and its relation to the eigenvalue. By disentangling the influence of system size, graph component size distribution, mean degree of nodes, network motifs, and degeneracy, we provide a comprehensive understanding of this system. We compare our findings to ordered graphs with the same mean degree and to one-dimensional tight-binding models. }

\section{Introduction}
The transport of charges in a semiconductor, the propagation of light in a crystal, and the dispersal of species in a patchy landscape are all affected by the presence of disorder. 
Starting from the seminal work by Anderson in 1958~\cite{andersonAbsenceDiffusionCertain1958}, it has become clear that a sufficiently strong disorder leads to a localization of eigenvectors of such systems.
Anderson considered a tight-binding model for electrons with random on-site potential. Subsequent work generalized these findings to tight-binding models with disorder also in the hopping terms, and to other wave equations that can be mapped on the same type of eigenvalue problem. 
Generally, all states are localized in one dimension~\cite{delyonOnedimensionalWaveEquations1983}, with two being the critical dimension for a disorder-induced metal-insulator phase transition~\cite{abrahamsScalingTheoryLocalization1979}. 

When site and bond disorder are correlated so that the Hamiltonian of the Anderson model becomes a Laplace matrix where the sum of all entries in a row vanishes, there is a conserved quantity, leading to system-spanning eigenvectors in finite systems even in the presence of strong disorder and even in one dimension~\cite{hirotsugamatsudaLocalizationNormalModes1970,schaeferScalingBehaviourLocalised2024}. Such models apply, for instance, to classical harmonic chains and to the diffusion of a substance or a biological population through a set of coupled sites. For one-dimensional systems, it was shown that the number of system-spanning eigenvectors scales as $1/\sqrt{N}$ with $N$ being the number of sites~\cite{hirotsugamatsudaLocalizationNormalModes1970, dominguez-adameDelocalizedVibrationsClassical1993,paytonDynamicsDistortedHarmonic1967,p.deanVibrationsTwodimensionalDisordered,schaeferScalingBehaviourLocalised2024}.
In the limit of infinitely large systems, this is a vanishing fraction of all modes, so that even in this case (almost) all modes are localized in 1D. 

More recently, research on the localization of eigenvectors was extended to complex networks. Complex networks incorporate disorder in their connection pattern and have "localized" modes that are concentrated on a small number of network nodes, in addition to delocalized modes. The structure of a complex network is characterized by its adjacency matrix $A_{ij}$, with $A_{ij}$ being 1 if the nodes $i$ and $j$ are connected and 0 otherwise. Evaluating the spectrum of the adjacency matrix of a network yields information on its structure~\cite{blackwellSpectraAdjacencyMatrices, spielmanSpectralAlgebraicGraph}. The spectrum of the Laplace matrix is relevant when studying the diffusion of a substance on the network or when studying the synchronizability of a set of coupled oscillators, the couplings being represented by the edges of the network~\cite{vanmieghemGraphSpectraComplex2023,diazSynchronizationInRandomGeometricGraphs2009}. Laplacian spectra are also relevant for the linear stability analysis of dynamical processes on networks (see, for example,~\cite{brechtelMasterStabilityFunctions2018}). 
Often, such studies are based on Erd\H{o}s-Rényi and Barabási-Albert networks~\cite{mcgrawLaplacianSpectraDiagnostic2008,zhuLocalizationsComplexNetworks2008}. However, these two types of network do not include the concept of spatial proximity. 
In contrast, many realistic networks, such as transportation networks, wireless networks, social networks, or ecological networks, are embedded in space (see~\cite{barthelemySpatialNetworks2011a} for a review of spatial networks). 
A generic model for networks embedded in space is given by random geometric graphs (RGGs)~\cite{penroseRandomGeometricGraphs2003,dallRandomGeometricGraphs2002,dettmannSymmetricMotifs2017,Dettmann_2017}. Here, nodes are randomly distributed in space, and pairs of nodes with a distance smaller than a cut-off (connectivity radius) are connected.
In one dimension, RGGs are a special case of interval graphs if the length of the intervals is constant, or circular-arc graphs for periodic boundary conditions and constant arc lengths.

The local connection rules of RGGs lead to motifs, that is, recurrent substructures of the network~\cite{nybergMesoscopicStructures2015}.
A subset of those motifs are orbits that result from two nodes sharing the same neighborhood. Orbits lead to eigenvectors that are fully localized on the nodes of the orbits and have the value zero elsewhere~\cite{nybergLaplacianSpectraRandom2014}. The associated eigenvalues of the Laplacian are positive integers and the associated eigenvalues of the adjacency matrix are -1 and 0, depending on whether the nodes sharing the same neighborhood are connected or not (type-I and type-II orbits, respectively).
Nyberg et al.~have found a closed mathematical expression for the number of type-I orbits in 1D RGGs with periodic boundary conditions~\cite{nybergLaplacianSpectraRandom2014,nybergMesoscopicStructures2015}. From this they conclude that approximately one third of the eigenvalues stem from type-I orbits in the limit of large $N$ and fixed $r$. 
The same holds in the thermodynamic limit when the mean degree is fixed and much larger than two.
Hamidouche et al.~\cite{hamidoucheNormalizedLaplacianSpectra2023} compared the spectra of normalized Laplacians (which are rescaled such that the diagonal elements are 1) with those of a lattice model where all nodes have the same number of neighbors, and they found upper bounds for the deviation between the two in the limit of large node numbers. These bounds decrease with increasing mean number of neighbors per node.  Similar results were obtained when the adjacency matrices were used instead of the normalized Laplacians \cite{hamidoucheSpectralAnalysisAdjacency2019a}.  The moments of the eigenvalue distribution of the adjacency matrix can be expressed in terms of the node density and connectivity radius, which in turn affects spreading processes, for instance of viruses~\cite{preciadoSpectralAnalysisVirus2009}.

In this work, we complement the cited studies of one-dimensional RGGs by an extensive investigation of the spectra of their adjacency and Laplace matrix. We focus on the participation ratio as a measure of localization and investigate its probability distribution and the relation between the participation ratios and the eigenvalues.  Although the Laplace and adjacency matrix encode the same information about the structure of the RGG, the localization properties of their eigenmodes are qualitatively different due to the conservation law encoded in the Laplace matrix.  By disentangling the influence of system size, mean degree, graph components, and orbits, we are able to provide an intuitive understanding of many observed features of the data. Comparison with  ordered graphs where each node has the same number of neighbors, and  
with the 1D tight-binding model yields additional insights.

\section{Models}
\begin{figure}[H]
\centering
\begin{tikzpicture}
    \draw[thin,gray] (0,0) -- (4,0);
    
    \filldraw[black] (0.5,0) circle (3pt); 
    \filldraw[black] (2,0) circle (3pt); 
    \filldraw[black] (1.2,0) circle (3pt);
    \filldraw[black] (3.5,0) circle (3pt);
    \draw[thick] (0.5,0) -- (1.2,0);
    \draw[thick] (1.2,0) -- (2,0);
    \draw[thick] (0.5,0) .. controls (1,0.8) and (3,0.8) .. (3.5,0);
    
    \node[below, gray, yshift=-5pt] at (0,0) {0};
    \node[below, gray, yshift=-5pt] at (4,0) {1};
    \draw[thin, gray] (0,0.1) -- (0,-0.1);
    \draw[thin, gray] (4,0.1) -- (4,-0.1);

    \draw[->, thick] (5,0) -- (6,0);

    \draw[thin,gray] (7,0) -- (11,0);
    \draw[thin, gray] (7,0.1) -- (7,-0.1);
    \draw[thin, gray] (11,0.1) -- (11,-0.1);
    
    \filldraw[black] (8,0) circle (3pt); 
    \filldraw[black] (9.5,0) circle (3pt); 
    \filldraw[black] (8.7,0) circle (3pt);
    \filldraw[black] (7,0) circle (3pt);
    \draw[thick] (8,0) -- (8.7,0);
    \draw[thick] (8.7,0) -- (9.5,0);
    \draw[thick] (7,0) -- (8,0);
    
    \node[below, gray, yshift=-5pt] at (7,0) {0};
    \node[below, gray, yshift=-5pt] at (11,0) {1};

\end{tikzpicture}

\caption{A sample graph visualizing the periodic boundary conditions and the shifting of the node positions such that the first node of one of the components is at 0, with a  all its adjacent nodes having larger position values - unless there are no consecutive nodes further than $r$ apart, in which case the graph has no gaps. This reordering leads to the adjacency matrix being block diagonal.}
\label{fig:model}
\end{figure}
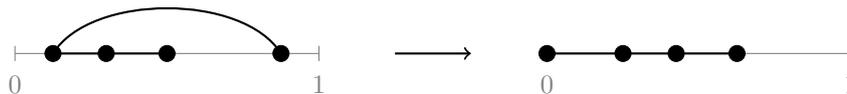

We generate one-dimensional random geometric graphs (RGG) of $N$ nodes by drawing $N$ random values from a uniform distribution on the interval $[0,1)$, which serve as coordinates of the nodes. Two nodes are connected if their distance is smaller than the connectivity radius, $r$. In the following, we assume periodic boundary conditions, that is, we distribute the nodes on a circle with circumference equal to 1.
There are $N!$ ways to label the $N$  nodes of the graph. We choose the labeling such that it preserves the ordering of the positions of the nodes. Furthermore, we shift the node positions so that the first node of one of the network components is at position 0, as 
illustrated in Fig.~\ref{fig:model}.
This has the effect that the adjacency matrix and the Laplace matrix become block diagonal.
The connectivity of the graph is determined by the parameter $z\coloneqq Nr$~\cite{nybergLaplacianSpectraRandom2014}, with the mean degree $\langle k\rangle$ being equal to $2z$. Smaller $z$ result in less connected graphs.

We consider two versions of the model, the Laplace model (LM) and the adjacency model (AM), i.e.,  we evaluate the spectra of the Laplace and the adjacency matrix (here: the negative adjacency matrix). Both of these spectra can be related to dynamical processes by considering the tight-binding Schroedinger equation
\begin{align}\label{LM}
    E\psi_n = \sum_{m\in\text{N(n)}}t_{nm}(\psi_n-\psi_m)
\end{align}
for the LM and
\begin{align}\label{AM}
    E\psi_n = -\sum_{m\in\text{N(n)}}t_{nm}\psi_m
\end{align}
for the AM, where $N(n)$ denotes the set of neighboring nodes of node $n$, $t_{nm}$ is the transition rate from node $m$ to node $n$ (and vice versa), $\psi_n$ is the amplitude of the wavefunction on node $n$, and $E$ is the total energy of the system. The LM is a model that has a conserved quantity $\sum_n \psi_n $. It applies, for instance, to a classical system where a substance with local concentration $\psi_n$ diffuses between connected sites. The full time-dependent dynamics of such a diffusion system corresponds formally to a Schr\"odinger equation with imaginary time, i.e., to
\begin{align}\label{LM_timedependent}
    \frac{\partial}{dt}\Psi_n(t) = \sum_{m\in\text{N(n)}}t_{nm}(\Psi_n(t)-\Psi_m(t))
\end{align}
and results in Eq.~\eqref{LM} when the ansatz $\Psi_n(t) = e^{-\lambda t}\psi_n$ is made. In this case, $E$ corresponds to the relaxation constant $\lambda$. Therefore, we will obtain only nonnegative values of $E$ for the LM.
The LM is an RGG analogue of the diffusion model (DM) on 1D lattices with random bonds $t_{ij}$ and with on-site energies $\epsilon_i$ such that they are equal to the negative sum of the coupling terms, $\epsilon_i = - \sum_j{t_{ij}}$ with $j$ being the nearest neighbors of $i$ \cite{schaeferScalingBehaviourLocalised2024}. The conservation law of these models leads to system-spanning modes, since an imbalance in the distribution of the diffusing substance can only be equilibrated by transport over the scale of the imbalance. When Eq.~\eqref{LM} is written in matrix form, the matrix is the Laplace matrix of the RGG, for which the sum of all elements in a row is 0.  When Eq.~\eqref{AM} is written in matrix form, the matrix is minus the adjacency matrix of the RGG. In contrast to the LM, the AM has no conserved quantities, allowing for a short-distance relaxation of a disequilibrium.

\begin{figure}[H]
\begin{subfigure}[b]{0.5\textwidth}
    \centering
    \includegraphics[width=\textwidth]{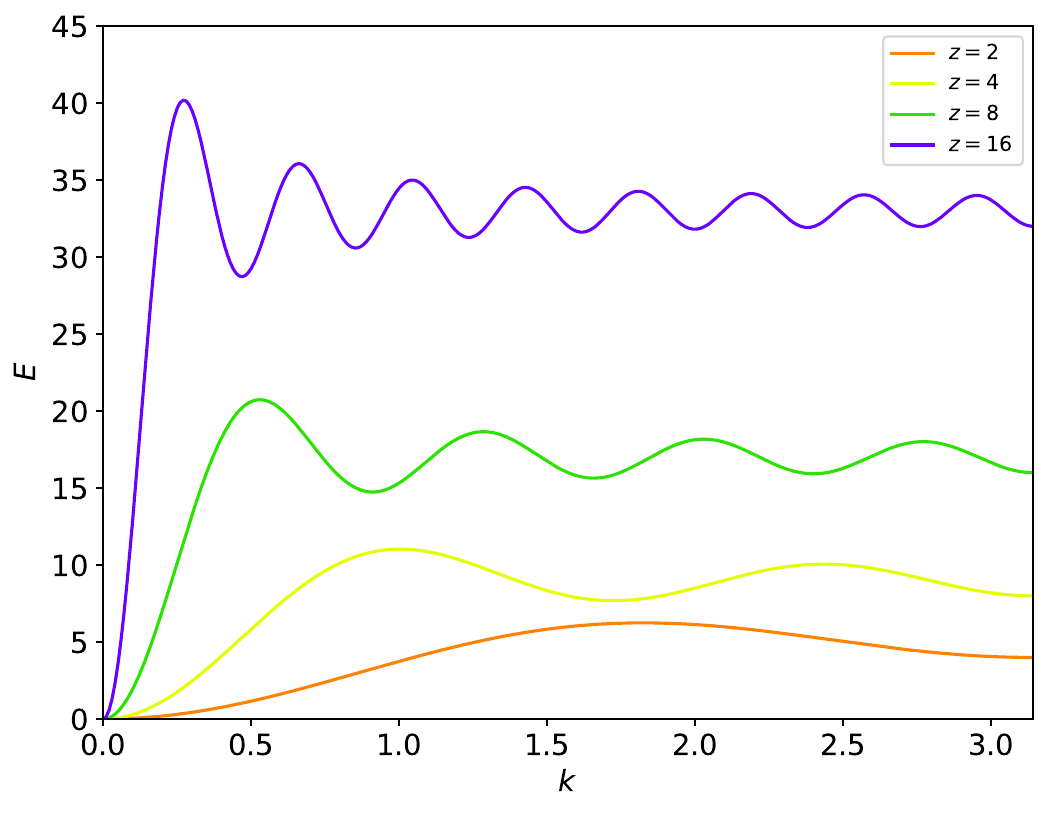}
\end{subfigure}
\begin{subfigure}[b]{0.5\textwidth}
    \centering
    \includegraphics[width=\textwidth]{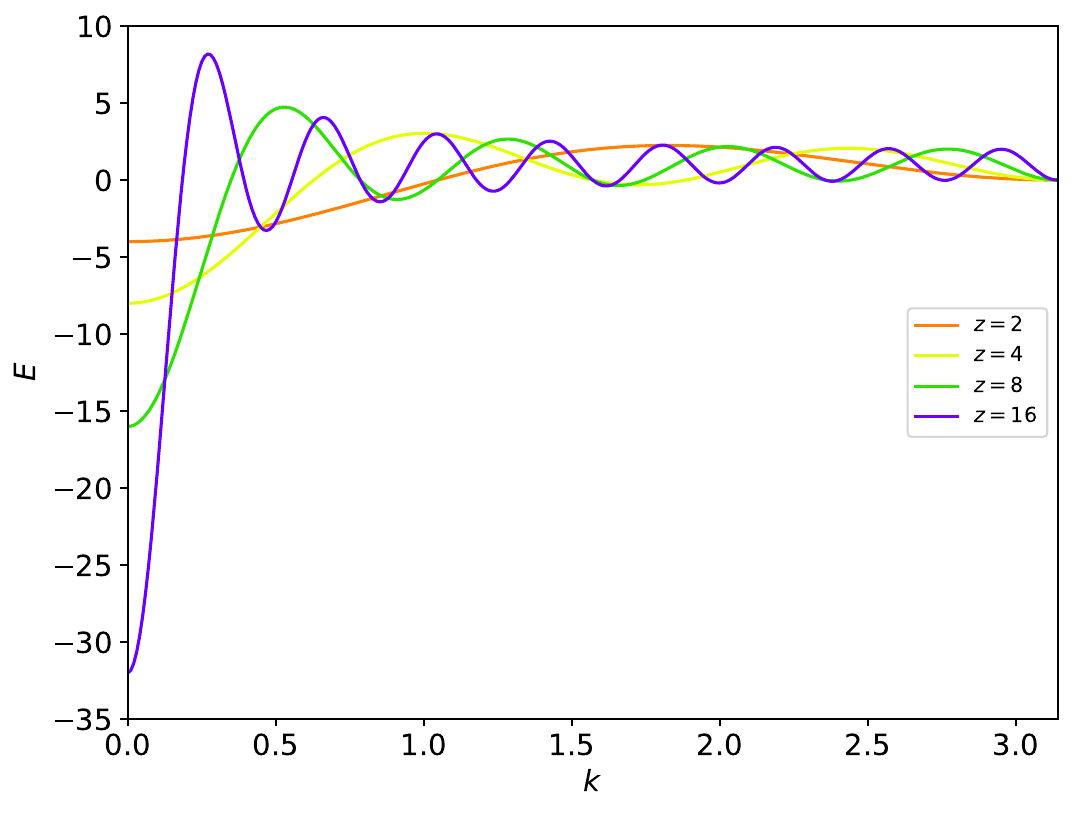}
\end{subfigure}
\caption{Dispersion relation of the Laplace model (left) and the adjacency model (right) for different values of $z$ for a regular lattice where each site has exactly $2z$ neighbors.}
\label{fig:dispersion_relation}
\end{figure}

On a regular 1D lattice, where each node has exactly $2z$ neighbors, the dispersion relation can be evaluated analytically and is given by
\begin{equation}
    E(k) = 4\sum_{n=1}^z \sin^2\left(n\frac{k}{2}\right) \label{eq:dispersionrelationLM}
\end{equation}
for the LM and 
\begin{align}
    E(k) = -2\sum_{n=1}^z \cos(nk)
\end{align}
for the AM.
Fig.~\ref{fig:dispersion_relation} depicts these dispersion relations. They differ by an energy shift of $2z$, as is evident from comparison of Eqs.~\eqref{LM} and \eqref{AM}. For $z\geq2$, the dispersion relation is not invertible on the interval $[0,\pi]$, so that there exist $z$ modes that have the same energy, resulting in several terms that contribute to the density of states at a given value of $E$. The typical energy value $2z$ of the Laplace model will manifest itself below when the Laplace spectrum of the RGG is evaluated, and the large weight at negative energies (compared to positive energies) of the AM will show below when the adjacency spectrum of the RGG is evaluated. 

\section{Methods}       
In order to quantify the localization behavior of the models, we calculate the eigenvectors and associated energies of Eqs.~\eqref{LM} and \eqref{AM} by diagonalizing the Hamiltonian using
the Julia library \texttt{LinearAlgebra}. We then determine their participation ratio $P$ and evaluate the probability distribution of $P$ and the relation between $P$ and the energy $E$. The features found in these data will be explained by considering the component size distribution of the RGGs, the orbits, the density of states (DOS), and by comparing to the model without disorder and to the lattice model with random couplings.

\subsection{Participation ratio}
The participation ratio is defined as
\begin{align}
    P=\frac{\left(\sum_n^N|\psi_n|^2\right)^2}{\sum_n^N|\psi_n|^4}~,
\end{align}
where the numerator is 1 when the wave function is normalised.
This definition has the properties we require for a localization measure.
That is,  $P=1$ for a state that sits at only one site, and  $P=N$ for a state with the same amplitude at every site. A special case, which we will encounter later, is the participation ratio of a sine function, for which we have  $P=2N/3$ in one dimension. For large $N$, the evaluation of the probability distribution of $P$ requires large computational resources as all $N$ eigenvectors of the system must be determined. Therefore, we limit our study mostly to system sizes up to $N=10^4$.

\subsection{Component size distribution}
The average distance between two neighboring nodes is $1/N$ where $N$ is the number of nodes. Due to the random placement of the nodes, the distribution of their nearest-neighbor distances $x$ is the exponential distribution $p(x)=N e^{-xN}$. The probability that this distance is larger than the connectivity radius, $r$ is  $P(x>r)=e^{-rN}$. The size distribution of the components $P(n)$ is given by the probability that $n$ consecutive nodes each have a distance smaller than $r$ and the next node has a distance larger than $r$,  
\begin{align}
    P(n)&=\left(1-e^{-rN}\right)^{n-1}e^{-rN}\\
    &= \frac{1}{e^{rN}-1} e^{n\ln(1-e^{-rN})}.
\end{align}
This is an exponential distribution in $n$. For $z=2,4,8,16$ (i.e. $r=2/N, 4/N, 8/N, 16/N$), we obtain approximately $P(n) \sim e^{-n/7}$, $e^{-n/54}$, $e^{-n/2980}$, $e^{-n/889000}, e^{-n/(7.9*10^{13})}$. For the system sizes $N \le 10000$ used in our simulations, the RGGs are composed of several unconnected components for $z=2$ and 4. For $z=8$, the RGG might have a system-spanning component, and for $z\ge16$, most of the RGG realizations span the entire system. The properties of the eigenvectors of the system will therefore change qualitatively with $z$, since the average size of the components on which the modes are located changes so strongly with $z$. 

\subsection{Orbits}
In contrast to lattice graphs, RGGs contain so-called orbits. These are motifs, that is, recurring substructures, in which two nodes share the same neighborhood, leading to eigenvectors that are fully localized on these nodes. 
In the spectrum of the Laplace matrix, such motifs appear as integer eigenvalues and contribute to its discrete part. These eigenvalues are directly related to the degree $k$ of the nodes (Eq.~\eqref{eq:orbits}). If the nodes themselves are not connected, the eigenvalue is their degree. If they are connected, the eigenvalue is their degree plus 1. The corresponding eigenvector $\vec v$ has the value 1 on one node (node 1) and -1 on the other (node 2), and 0 on all remaining nodes:
\begin{align}\label{eq:orbits}
    L\Vec{v} &= L(\Vec{e}_1-\Vec{e}_2) \nonumber\\
    &= \sum_{i,j=1}^N \Vec{e}_i L_{ij} v_{j} \nonumber\\\
    &=  \sum_{i,j=1}^N \Vec{e}_i  L_{ij}(\delta_{j1}- \delta_{j2})\nonumber\\\
    &= \sum_{i=1}^N \Vec{e}_i (L_{i1} - L_{i2})\nonumber\\\
    &= (k-L_{12})(\Vec{e}_1-\Vec{e}_2)~.
\end{align}
If $n>2$ nodes share the same neighborhood, they all have the same degree, and the above considerations apply to each subset of 2 nodes. Since it is impossible that more than two nodes of a one-dimensional RGG share the same neighborhood but are not connected among each other, the eigenvalues of the Laplace matrix must be $k+1$ in this case. Linear combinations of the eigenvectors pairs of nodes are also eigenvectors of the eigenvalue $k+1$, with the degeneracy being $n-1$. 

The (negative) adjacency matrix has the same eigenvectors, and the eigenvalues are -1 if the nodes of the motif are connected and 0 otherwise. 

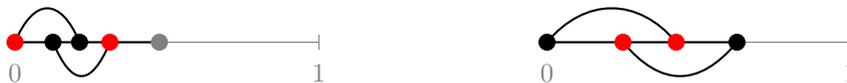
\begin{figure}[H]
\centering
\begin{tikzpicture}
    \draw[thin,gray] (0,0) -- (4,0);
    \draw[thin, gray] (0,0.1) -- (0,-0.1);
    \draw[thin, gray] (4,0.1) -- (4,-0.1);

    \draw[thick] (0,0) -- (0.5,0);
    \draw[thick] (0.5,0) -- (0.85,0);
    \draw[thick] (0.85,0) -- (1.25,0);
    \draw[thick] (1.25,0) -- (1.9,0);

    \draw[thick] (0,0) .. controls (0.25,0.6) and (0.6,0.6) .. (0.85,0);
    \draw[thick] (0.5,0) .. controls (0.75,-0.6) and (1,-0.6) .. (1.25,0);
    
    \filldraw[red] (0,0) circle (3pt);
    \filldraw[black] (0.5,0) circle (3pt); 
    \filldraw[black] (0.85,0) circle (3pt);
    \filldraw[red] (1.25,0) circle (3pt);
    \filldraw[gray] (1.9,0) circle (3pt);
    
    \node[below, gray, yshift=-5pt] at (0,0) {0};
    \node[below, gray, yshift=-5pt] at (4,0) {1};

    
    \draw[thin,gray] (7,0) -- (11,0);
    \draw[thin, gray] (7,0.1) -- (7,-0.1);
    \draw[thin, gray] (11,0.1) -- (11,-0.1);

    \draw[thick] (8,0) -- (8.7,0);
    \draw[thick] (8.7,0) -- (9.5,0);
    \draw[thick] (7,0) -- (8,0);

    \draw[thick] (7,0) .. controls (7.5,0.6) and (8.2,0.6) .. (8.7,0);
    \draw[thick] (8,0) .. controls (8.5,-0.6) and (9,-0.6) .. (9.5,0);
    
    \filldraw[red] (8,0) circle (3pt); 
    \filldraw[black] (9.5,0) circle (3pt); 
    \filldraw[red] (8.7,0) circle (3pt);
    \filldraw[black] (7,0) circle (3pt);   
    
    \node[below, gray, yshift=-5pt] at (7,0) {0};
    \node[below, gray, yshift=-5pt] at (11,0) {1};

\end{tikzpicture}

\caption{A sample graph visualizing Type-I (left) and Type-II orbits (right). In both cases, the black nodes have the same neighborhood (red nodes) but for Type-I orbits, the black nodes are also connected. The left configuration leads to the eigenvalue 4 and the right to the eigenvalue 2 of the Laplace matrix. In 1D RGGs Type-II orbits are always isolated.}
\label{fig:model_orbits}
\end{figure}

\subsection{Data handling}
The simulations were run in Julia 1.10.1. The algorithm used for the exact diagonalization is \texttt{eigen()} from the \texttt{LinearAlgebra} library.
The random positions of the nodes were generated with \texttt{rand()}, which returns pseudorandom numbers (Xoshiro256++) distributed uniformly in the interval $[0,1)$.
We sorted the nodes by position, with position zero coinciding with the first node of a component. This means that the Hamiltonian becomes a block diagonal matrix, and the eigenvectors produced by the algorithm are confined to a single component each, even when there is degeneracy. 

The data shown in the figures is based on the average of an ensemble of 1,000 realizations for the system sizes 625, 2500, and 10,000. Due to hardware limitations, only 8 and 12 realizations were simulated for $N=40,000$ for the DM and the AM, respectively.
If not stated otherwise, bins with less than five counts were excluded.

\section{Results}
\subsection{Density of states}
\begin{figure}[H]
        \centering
        \begin{subfigure}[b]{0.5\textwidth}
            \centering
            \includegraphics[width=\textwidth]{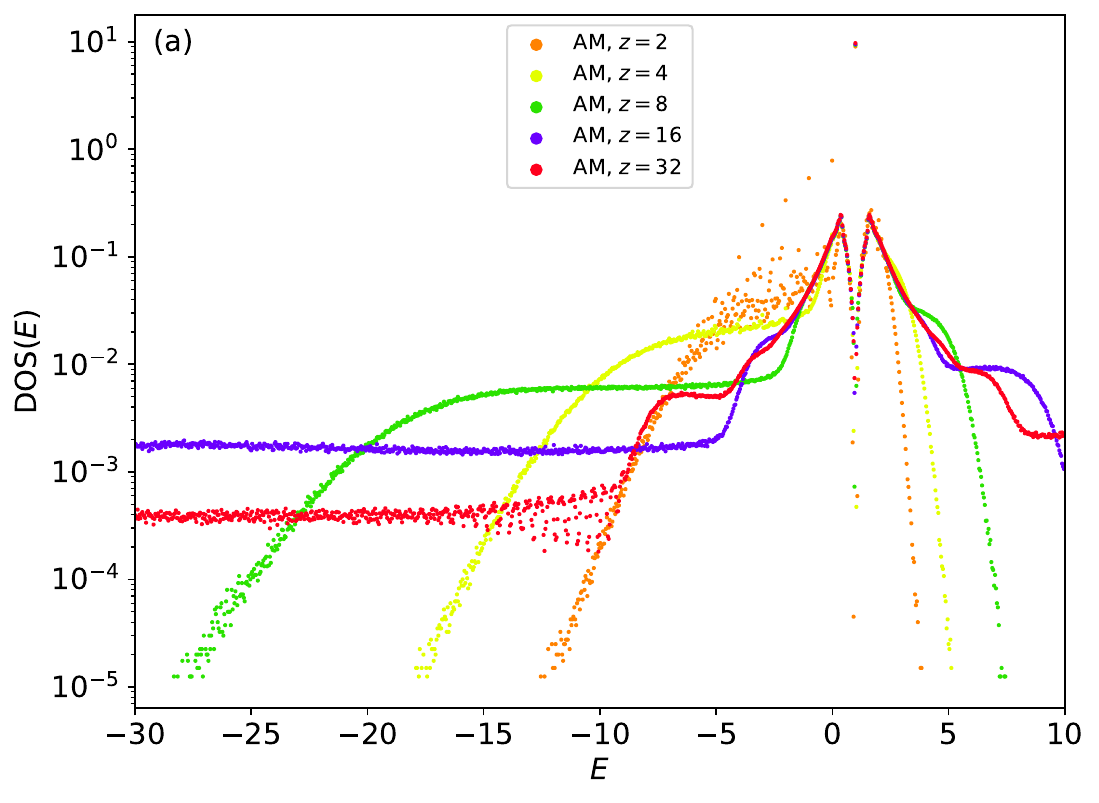} 
            \label{fig:subfig1}
        \end{subfigure}%
        \hfill
        \begin{subfigure}[b]{0.5\textwidth}
            \centering            \includegraphics[width=\textwidth]{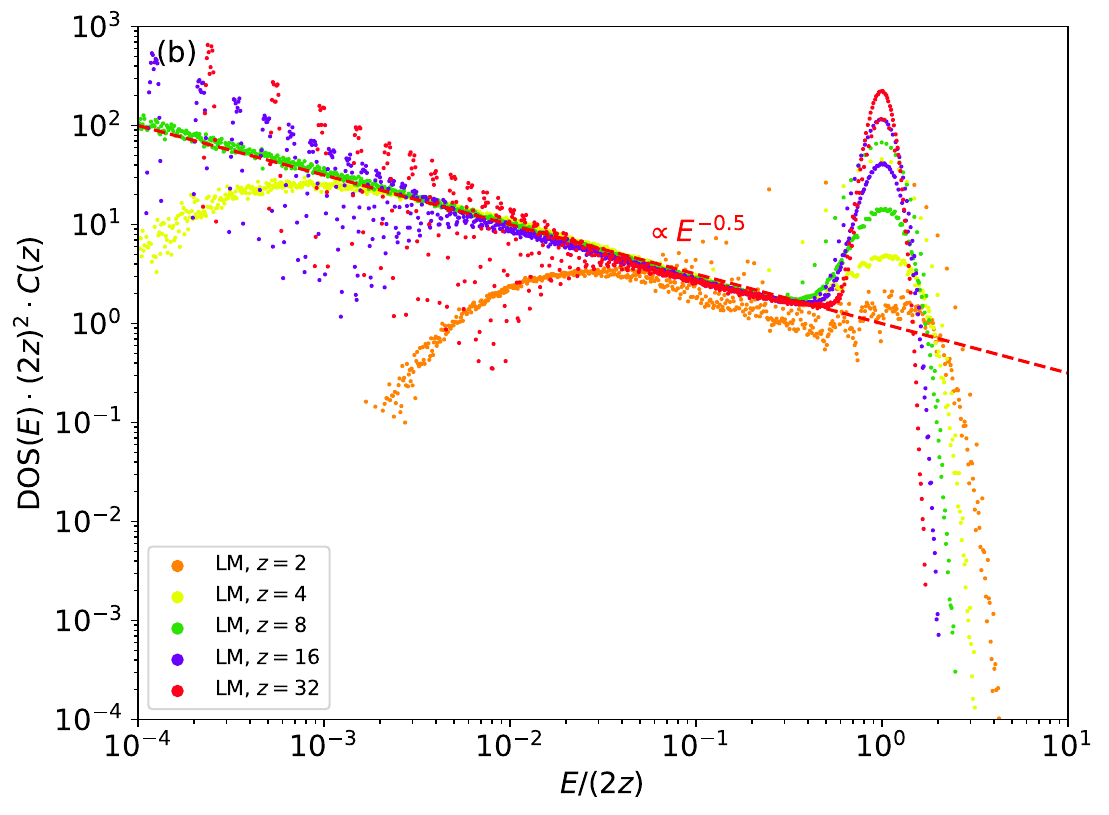} 
            \label{fig:subfig2}
        \end{subfigure}
        \centering
        \begin{subfigure}[b]{0.5\textwidth}
            \centering
        \includegraphics[width=\textwidth]{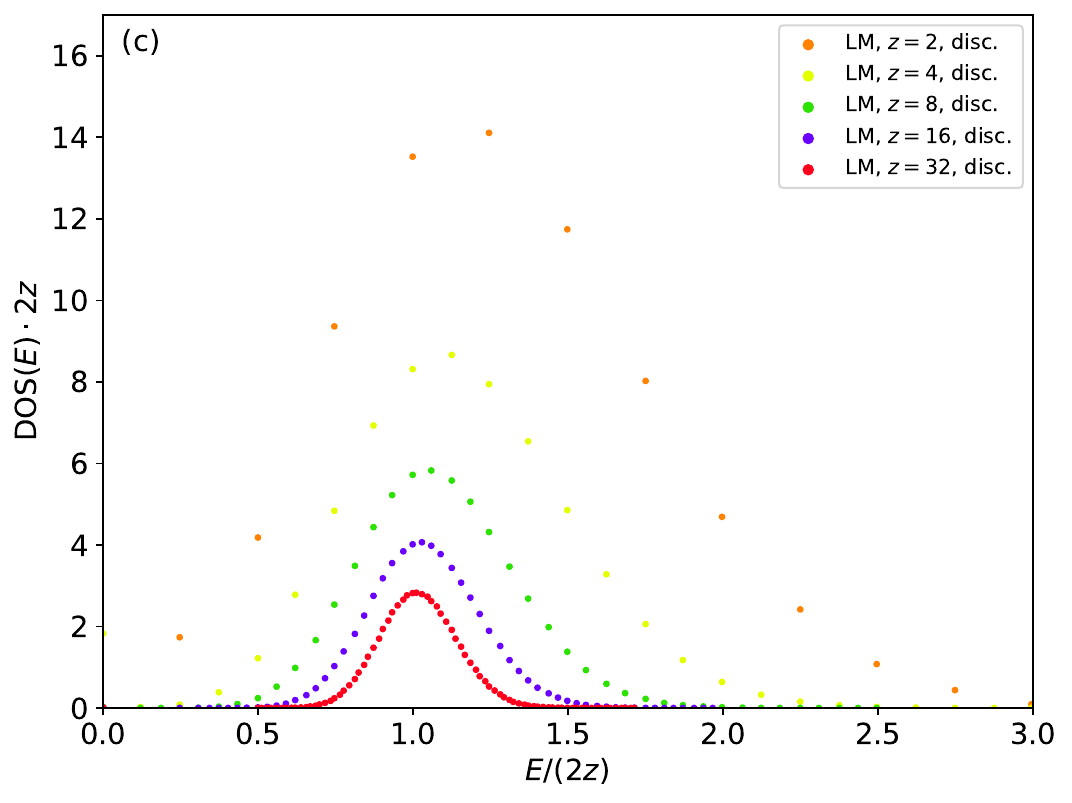} 
            \label{fig:subfig3}
        \end{subfigure}%
        \hfill
        \begin{subfigure}[b]{0.5\textwidth}
            \centering
            \includegraphics[width=\textwidth]{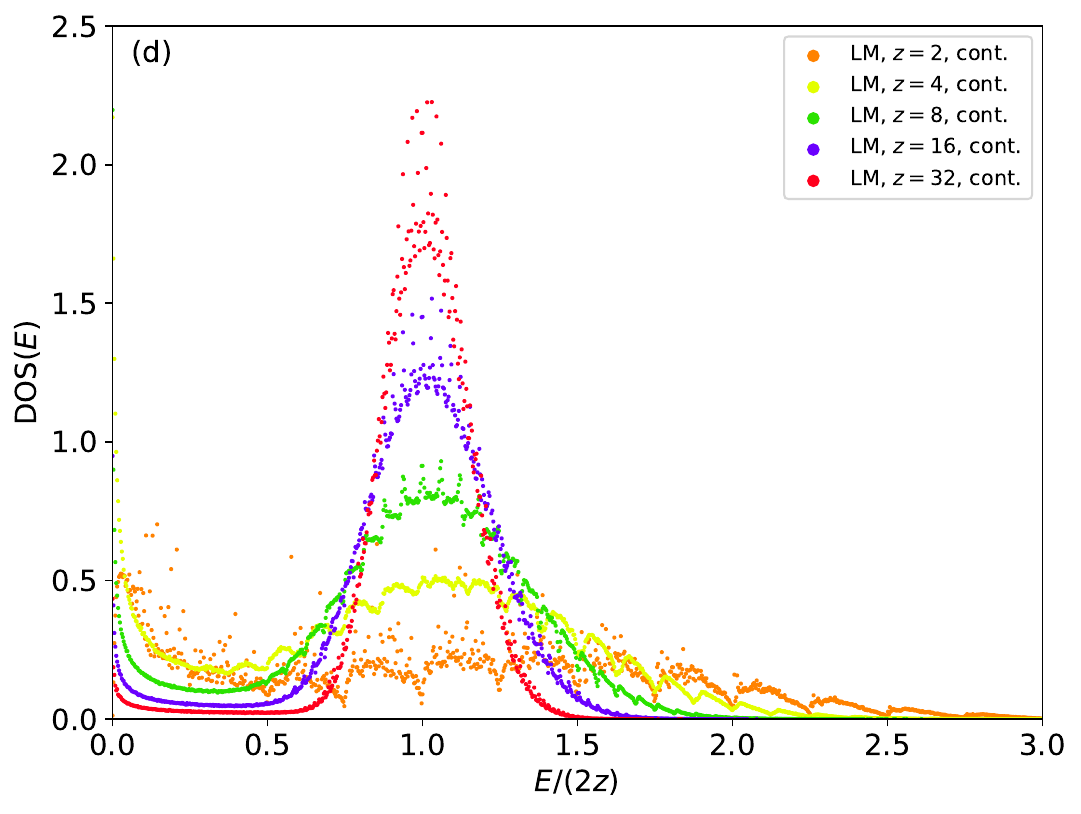}
            \label{fig:subfig4}
        \end{subfigure}

    \caption{Density of states  for the AM (a) and LM (b) for systems of size $10^4$. The $y$-axis of (b) is scaled by $(2z)^2\cdot C(z)$ where $C(z)=\sqrt{(1+1/z)(1+1/(2z))}$ (see~Eq.~\eqref{eq:DOS_correction}). Graphs (c) and (d) show for the LM the discrete and continuous part separately (i.e. the part for integer and non-integer energy values).}
    \label{fig:DOS}
\end{figure}

\begin{figure}[H]
        \centering
        \begin{subfigure}[b]{0.5\textwidth}
            \centering
            \includegraphics[width=\textwidth]{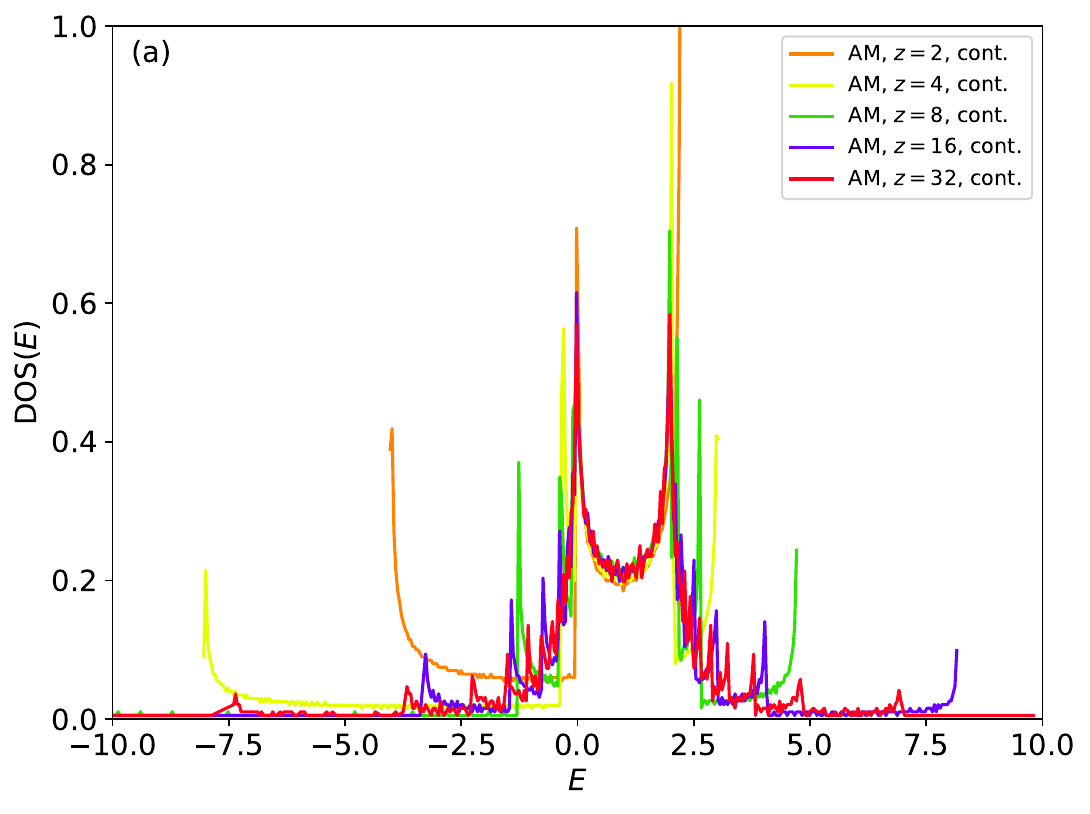} 
            \label{fig:subfig1}
        \end{subfigure}%
        \hfill
        \begin{subfigure}[b]{0.5\textwidth}
            \centering            \includegraphics[width=\textwidth]{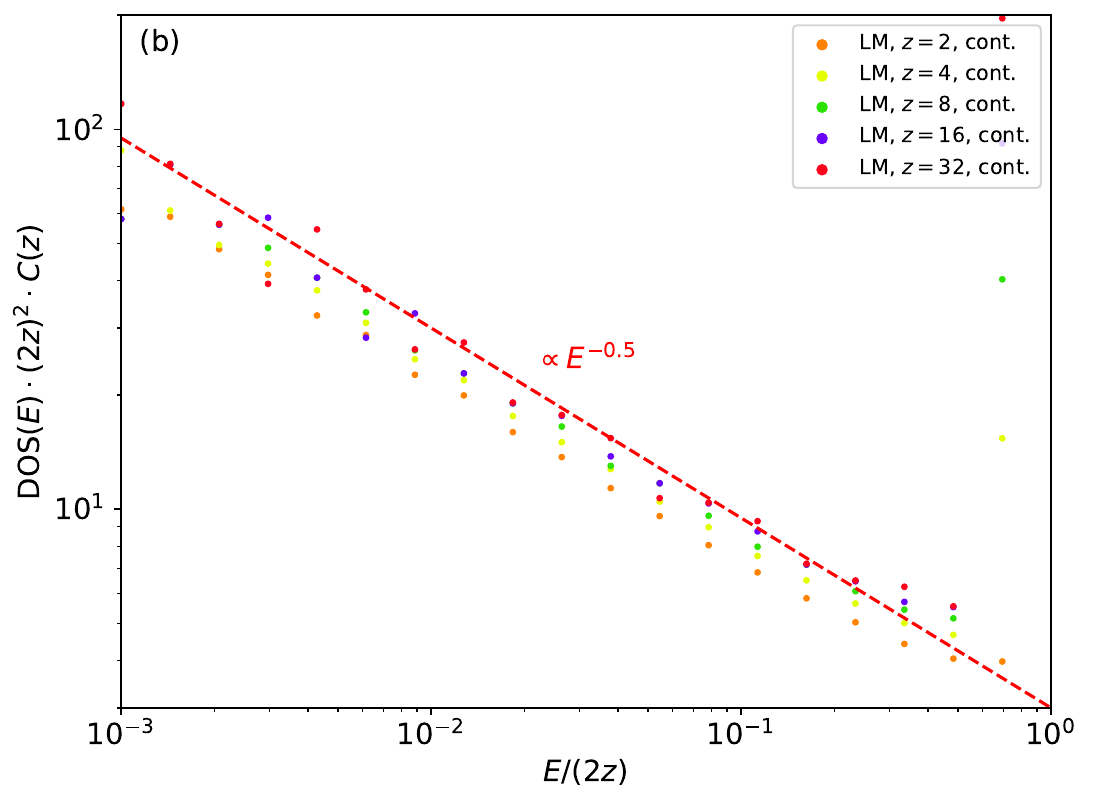} 
            \label{fig:subfig2}
        \end{subfigure}
    \caption{Density of states of a regular lattice, where each site has exactly $2z$ neighbors. For the LM, we chose logarithmic axes so that the power law $-1/2$ for small energies becomes visible. In fact, the DOS of the LM and AM are identical, apart from a shift of the energy values by $2z$, which follows directly from the corresponding shift for the dispersion relations (see Fig.~\ref{fig:dispersion_relation}).}
    \label{fig:DOSDGG}
    \end{figure}

In the spectrum of the AM (Fig.~\ref{fig:DOS}a),  there is a discrete peak resulting from the orbits at $E=1$. 
Around this discrete peak, the spectrum shows a funnel-shaped plunge, which must be due to points moving exactly to the value 1.
For higher absolute values of the energy, the DOS of the AM exhibits step-like features for $z \ge 8$, where the graph has system-spanning components. Compared with the ordered case where each node has exactly $2z$ neighbors (Fig.~\ref{fig:DOSDGG}), it is apparent that these steps occur at the band edges of the ordered system where the DOS diverges. Similarly, the flat parts of the DOS of the AM can be explained by comparison with the ordered case: in the vicinity of its local extrema, the function $E(k)$ is nonzero (see Fig.~\ref{fig:dispersion_relation}(a)), leading to a constant density of states in this energy region. Such connections between the DOS of the disordered and regular system were also observed in the 1D random coupling model~\cite{schaeferScalingBehaviourLocalised2024}. Beyond the maximum values of $|E|$ of the ordered model, the density of eigenstates of the RGG decreases rapidly.

In the LM, the spectrum has a continuous part and a discrete part, with the discrete peaks at integer values of $E$, corresponding to orbits in the RGG (Fig.~\ref{fig:DOS})~\cite{nybergMesoscopicStructures2015}. The continuous and the discrete part show a maximum around the mean degree $2z$, which determines the order of magnitude of the majority of eigenvalues. 
For small $E$, the density of states follows, for $z>2$, a power law $\sim 1/\sqrt E$ (red dashed line), just as for the ordered system with exactly $2z$ neighbors per site (see Fig.~\ref{fig:DOSDGG}b). These are the modes with large wavelength and thus a large extension. For $z=2$ the cutoff in component size is too small to show this power law. 
The dependence of the DOS on $E$ for small $E$ of the ordered system can be determined analytically from the dispersion relation \eqref{eq:dispersionrelationLM}:
We have
\begin{align}
    E(k) \simeq 4\sum_{n=1}^z \frac{n^2 k^2}{4} = k^2\frac{z(z+1)(2z+1)}{6} \, , \label{eq:dispersionrelationapprox}
\end{align}
and from the density $N/2\pi$ of $k$ values we obtain the density of $E$ values as
\begin{align}\label{eq:DOS_correction}
    \mathrm{DOS}(E) = \frac N{2\pi} \frac 1 {\mathrm{d}E/\mathrm{d}k}= \frac N{2\pi}\frac{3}{\sqrt{6Ez(z+1)(2z+1)}} \, .
\end{align}
This means that if we plot $\mathrm{DOS}(E)\cdot 4z^2\sqrt{(1+1/z)(1+1/2z)}$ versus $E/2z$, the curves for different $z$ coincide for small $E$. For $z=2$ and $4$, the cutoff in the component size distribution leads to a cutoff for the power law. For $z=16$ and $z=32$, we see a scattered set of data below the power law, which we ascribe to rare components of a size smaller than the system size. On the other hand, we see data above the power law, which are due to finite-size effects since the sine function modes must have at least one wavelength, and the corresponding energy is $E\simeq 5.9 \cdot 10^{-4}$ for $z=16$ and $E\simeq 4.5 \cdot 10^{-3}$ for $z=32$ (compare Eq.~\eqref{eq:dispersionrelationapprox}).

The continuous part of the density of states shows a series of bows, which is best visible for the smaller $z$ due to the scaling of the $x$-axis by $2z$. 
The minima between the bows occur at integer values of $E$, where the discrete spectrum has its data points (see Fig.~\ref{fig:DOS}). This seems to be the same effect as for the AM, where there is a funnel-shaped plunge around 1. Now, the discrete spectrum has points at all integer values of $E$, and thus there are funnels around each of them, resulting in the bows. 

Another striking feature of the DOS is the fuzzyness of the data, which is most pronounced for the main peak and for larger $z$. The data do not become sharper when the sample size is increased. This means that nearby energy values occur with different frequency. We explain this by the limited number of different connection patterns for subsets of several neighboring nodes. These connection patterns occur with different frequencies. Eigenvectors localized on such subsets have a discrete set of possible eigenvalues with their associated frequency of occurrence in the statistical ensemble. 

\subsection{Selected eigenvectors}
Since the Hamiltonian has block-diagonal form, all eigenvectors are confined to one component. However, when there is degeneracy of modes within one component, the algorithm gives linear combinations of these modes.

\begin{figure}[H]
    \centering
    \includegraphics[width=\linewidth]{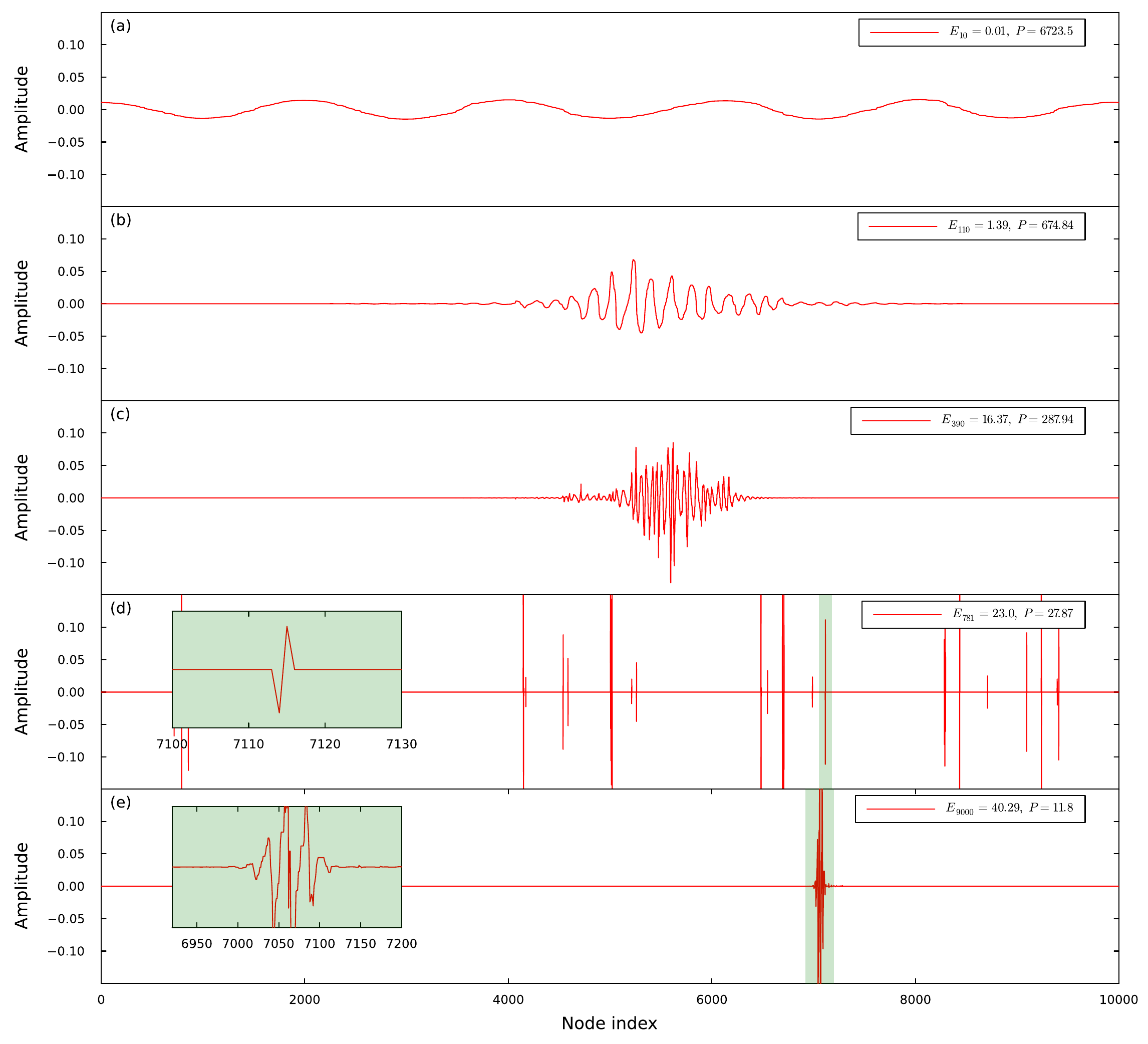}
    \caption{Selected eigenvectors of the LM for $z=16$ and $N=10^ {4}$: (a) system-spanning eigenvector with a large wavelength, (b) localized eigenvector for an energy in the minimum of the DOS between $E=0$ and $E=2z$ resembling a sine function with fluctuations of the amplitude due to disorder, (c) localized eigenvector for an energy also in this minimum but closer to the peak at $E=2z$, leading to a shorter wavelength, (d) eigenvector for an integer energy showing the localization on orbits, and (e) strongly localized eigenvector in the peak of the DOS. The insets show zooms into the shaded region.}
    \label{fig:EV_LM}
\end{figure}

\begin{figure}[H]
    \centering
    \includegraphics[width=\linewidth]{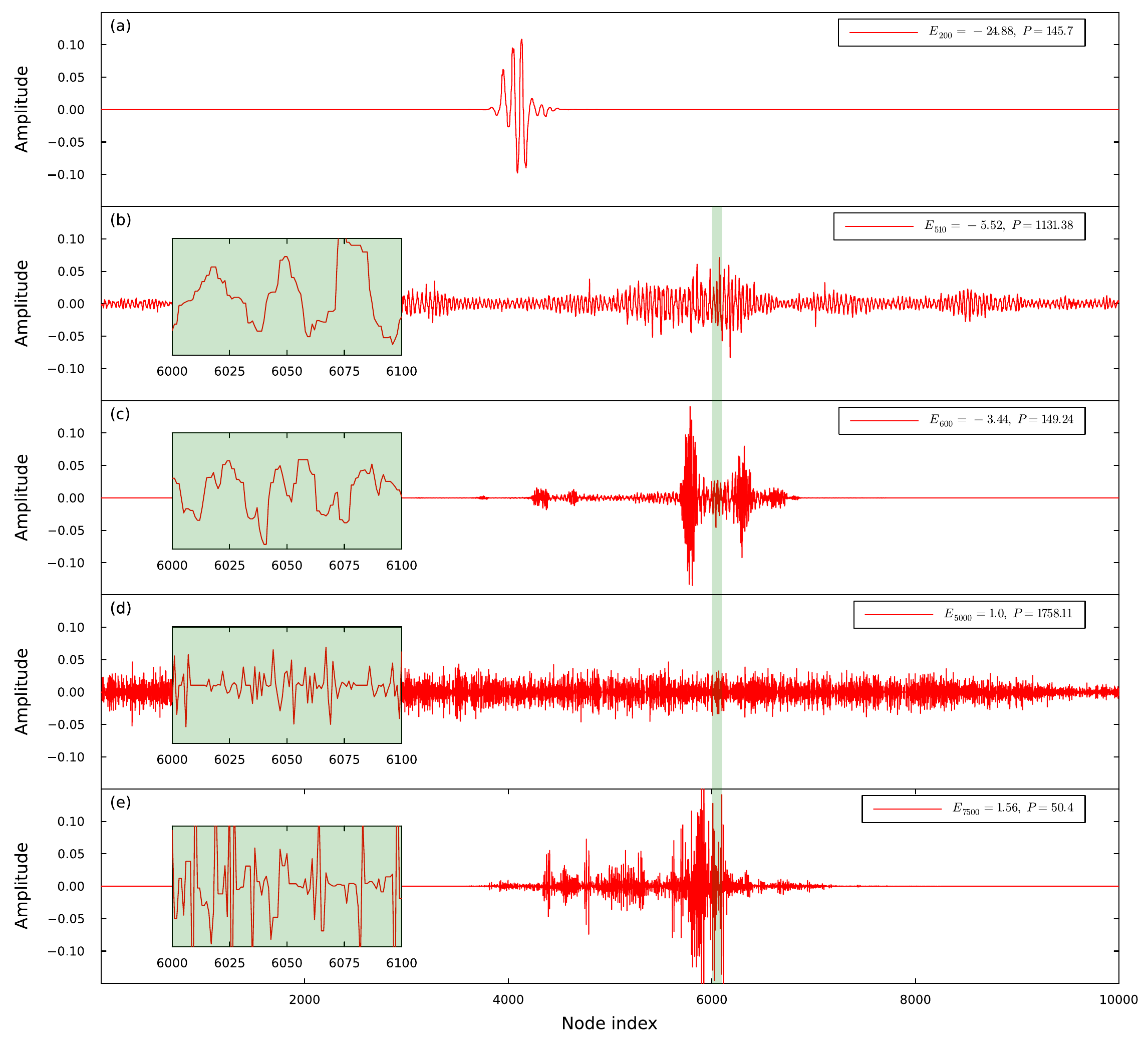}
    \caption{Selected eigenvectors of the AM for $z=16$ and $N=10^4$: (a) eigenvector close to $E=0$, (b) superposition of the degenerate eigenvalue $E=1$ due to orbits, (c) localized eigenvector close to $E=1$, and (d) localized eigenvector. }
    \label{fig:EV_AM}
\end{figure}

For the AM, all orbits have the same energy $E=1$, and therefore the degeneracy is greater, manifested in larger participation ratios (compare Fig.~\ref{fig:EV_AM}(d) with Fig.~\ref{fig:EV_LM}(d)). 

The modes belonging to the continuous spectrum change qualitatively as their energy changes. For the LM, the smallest energies belong to modes that resemble the sine functions of the ordered system. For larger energies, modes become increasingly localized; see Fig.~\ref{fig:EV_LM}(a-c),(e)). 

For the AM, the degree of localization changes in a nonmonotonic way with the energy, see Fig.~\ref{fig:EV_AM}. Negative eigenvalues with a large absolute value require the $\psi_n$ values of the sites that contribute to the eigenvector to preferably have the same sign, see Eq.~\eqref{AM}, leading to localized modes with a wavelength that is determined by the persistence length of the sign of $\psi_n$ (Fig.~\ref{fig:EV_AM}(a)). With increasing energy, the wave length becomes shorter (compare Fig.~\ref{fig:dispersion_relation}(b)) and the participation ratio becomes larger, with some eigenvectors extending over the entire system (Fig.~\ref{fig:EV_AM}(b)). As the eigenvalue $E=1$ is approached, the eigenvectors become increasingly localized (Fig.~\ref{fig:EV_AM}(c) and (e)). At $E=1$, the eigenvectors are superpositions of orbits, all of which have the same eigenvalue (Fig.~\ref{fig:EV_AM}(d)).

\begin{figure}[H]
    \centering
    \includegraphics[width=\linewidth]{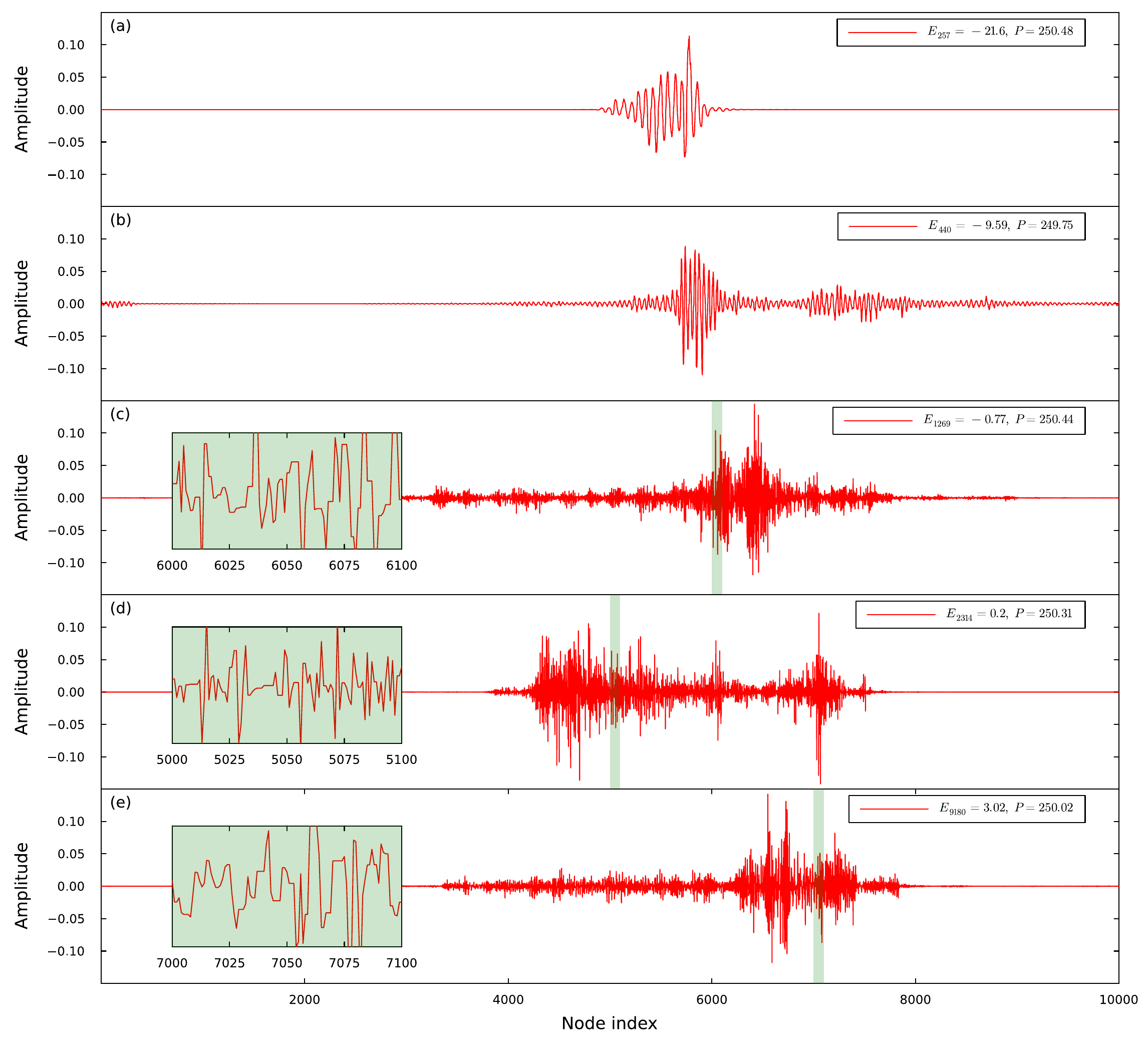}
    \caption{Selected eigenvectors of the AM for $z=16$ and $N=10^4$ whose participation ratio is close to 250, shown in ascending order of the energy.}
    \label{fig:EV_AM_same_P}
\end{figure}

This nonmonotonous behavior of the extension of eigenvectors with changing energy leads to the curious feature that eigenvectors of vastly different energies can have similar participation ratios. Fig.~\ref{fig:EV_AM_same_P} shows five eigenvectors with $P\approx 250$ but different energies. Again, modes with very negative energies have wavelengths that span several tens of sites, with sign changes becoming more frequent with increasing energy. Positive energies require frequent sign changes, see Eq.~\eqref{AM}.

\subsection{Probability distribution of the participation ratio}
\subsubsection{LM}
The distribution of the participation ratio (PPDF) changes in a nontrivial way with $z$ and system size, see Fig.~\ref{fig:PPDF_scaled}. The features of the distribution result from the interplay of the component size distribution, of the discrete and continuous part of the spectrum, of localized modes and system-spanning modes. Let us first consider the data for the LM. For $z=2$ and $z=4$, the PPDF has a cutoff much smaller than the system size, and this is due to the component size distribution having a cutoff much smaller than the system size. For $z=2$, the data points due to the orbits lie considerably above the other data, see the series of discrete orange points above the main curve in Fig.~\ref{fig:PPDF_scaled}(a). The maximum possible participation ratio is identical to the largest component size, unless degeneracy of energy eigenvalues leads to eigenvectors that cover several components. However, since we implemented the Hamiltonian in block diagonal form, the diagonalization algorithm yields only eigenvectors that are confined to one component. (What happens when no block-diagonal form is used is shown below in Sec.~\ref{sec:blockdiagonal}.)  For larger $z$, there exist system-spanning components and associated system-spanning eigenvectors that resemble the sine-function modes of the ordered system (see Fig.~\ref{fig:EV_LM}(a)), which manifest themselves in the final increase at the right end of the PPDFs. Apart from this peak, the PPDFs for $z=8,16,32$ show four additional features: At small participation ratios, there are localized modes, as can be seen from Fig.~\ref{fig:PPDF_scaled}(b), where the left part of the curve does not change any more for the larger system sizes. Next, there is a "nose" near $P=100$ that moves to smaller $P$ values with increasing $z$ and that vanishes when only the continuous part of the spectrum is considered (see the thinner curves marked with a small +). This nose must be due to orbits, as they dominate the discrete part of the spectrum. The nose moves to the right with increasing system size, indicating that several identical orbits that are located on the same component contribute to the same eigenvector, see Fig.~\ref{fig:EV_LM}(b). The third feature is a power-law behavior, which is best visible in the $z=8$ data in the interval $P=200$ to 3000. For the $z=16$ data, this power law becomes visible with increasing $N$, indicating that the cutoff of the power law depends on system size. The exponent of the power law is $-1.5$, which is the value that we observed for the regular 1D lattice for the "diffusion model"~\cite{schaeferScalingBehaviourLocalised2024}, where the Hamiltonian is a discretized Laplacian for a system with random nearest-neighbor couplings in the interval $[0,2]$ and corresponding on-site energies. This power law is due to modes that span many lattice sites and several wavelengths, but are still localized, see Fig.~\ref{fig:EV_LM}(b). We expect that for $z=32$ this power law would show up if the system size was made considerably larger. Our data show instead a pronounced hump at large P for the $z=32$ data, with the power law just starting to build up at the right-hand slope of this hump. This build-up of the power law at the right-hand side of the hump can be nicely seen in the $z=16$ data with increasing system size (Fig.~\ref{fig:PPDF_scaled}(b)). The hump itself, which is the fourth characteristic feature of the data, is due to there being a pronounced minimum left of the hump. The location of the minimum on the $P$ axis does not appear to depend on $N$, and it marks the upper boundary of the well-localized part that is independent of system size and where the energy is of the order of $2z$, compare Fig.~\ref{fig:EvP}(a) and (b). Around the energy $E=z$, the DOS shows a minimum, see Fig.~\ref{fig:DOS}(d), and this minimum shows also in the PPDF. The minimum in the DOS becomes more pronounced when $z$ is larger, explaining the greater height of the drop in the PPDF for increasing $z$.

\subsubsection{AM}
\begin{figure}[H]
    \centering
    \begin{minipage}{\textwidth}
        \centering
        \begin{subfigure}{0.5\textwidth}
            \centering
            \includegraphics[width=\linewidth]{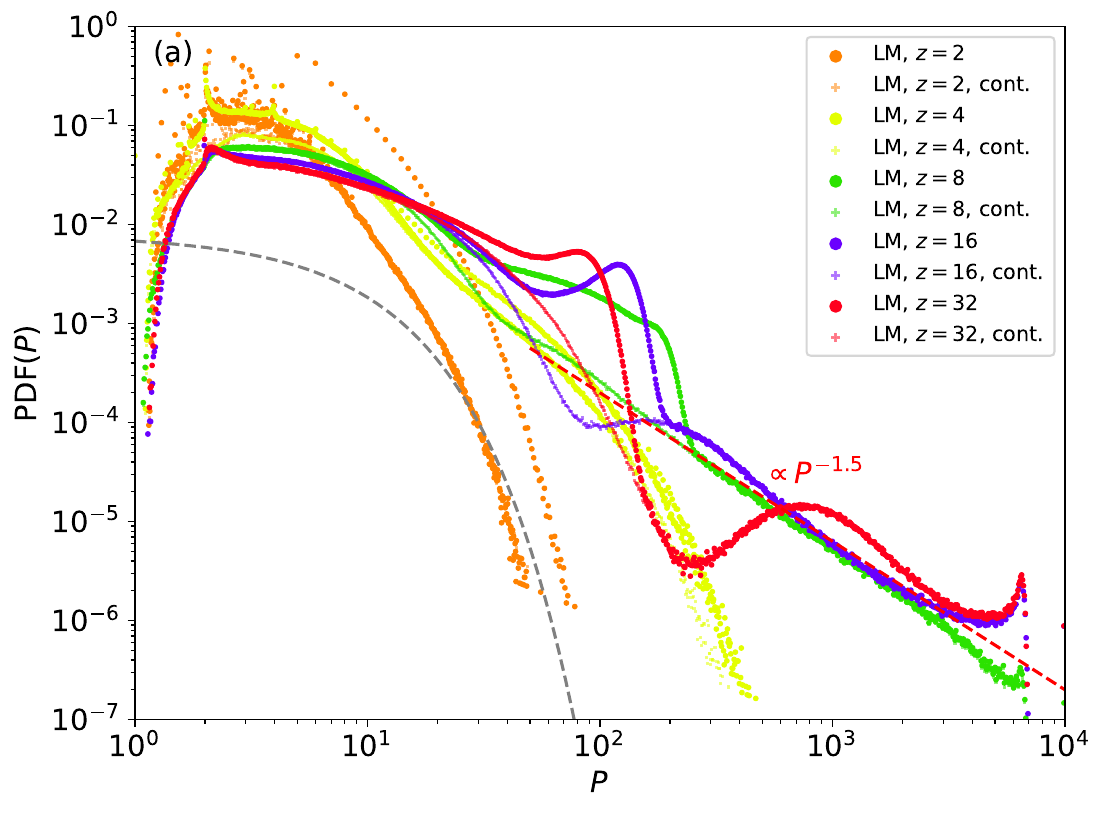} 
            \label{fig:subfig1}
        \end{subfigure}%
        \begin{subfigure}{0.5\textwidth}
            \centering
            \includegraphics[width=\linewidth]{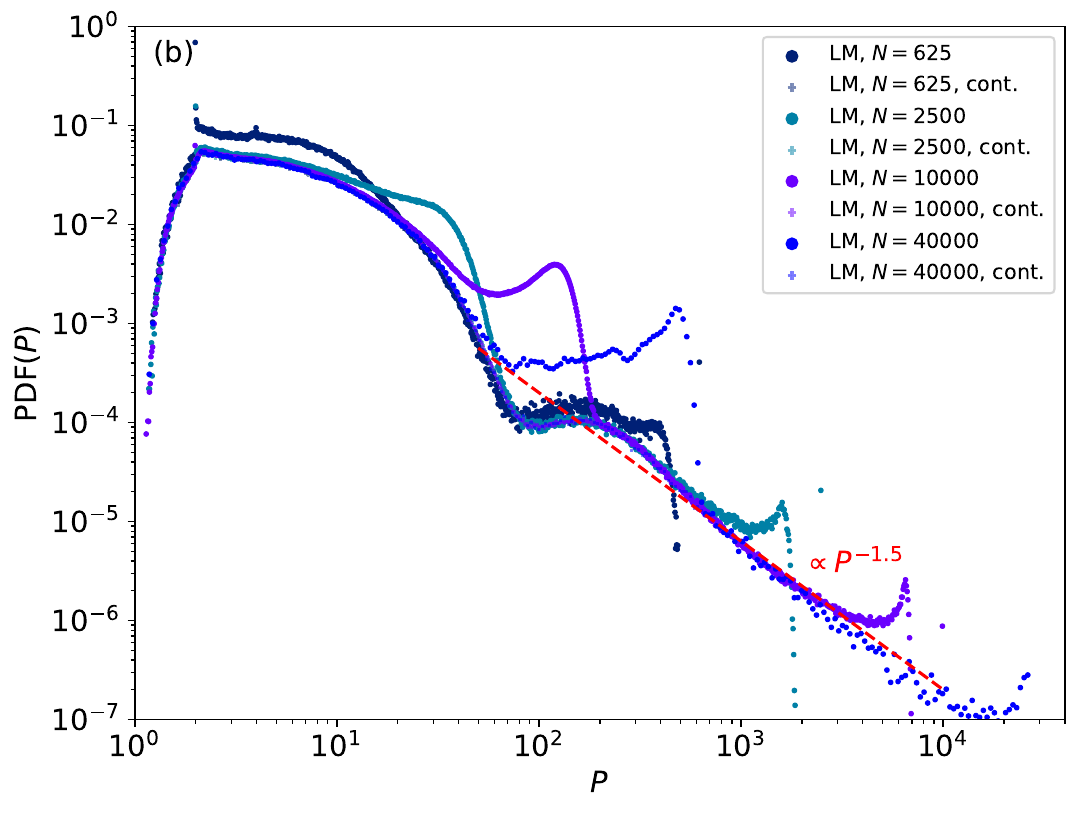} 
            \label{fig:subfig2}
        \end{subfigure}
    \end{minipage}
    \begin{minipage}{\textwidth}
        \centering
        \begin{subfigure}{0.5\textwidth}
            \centering
            \includegraphics[width=\linewidth]{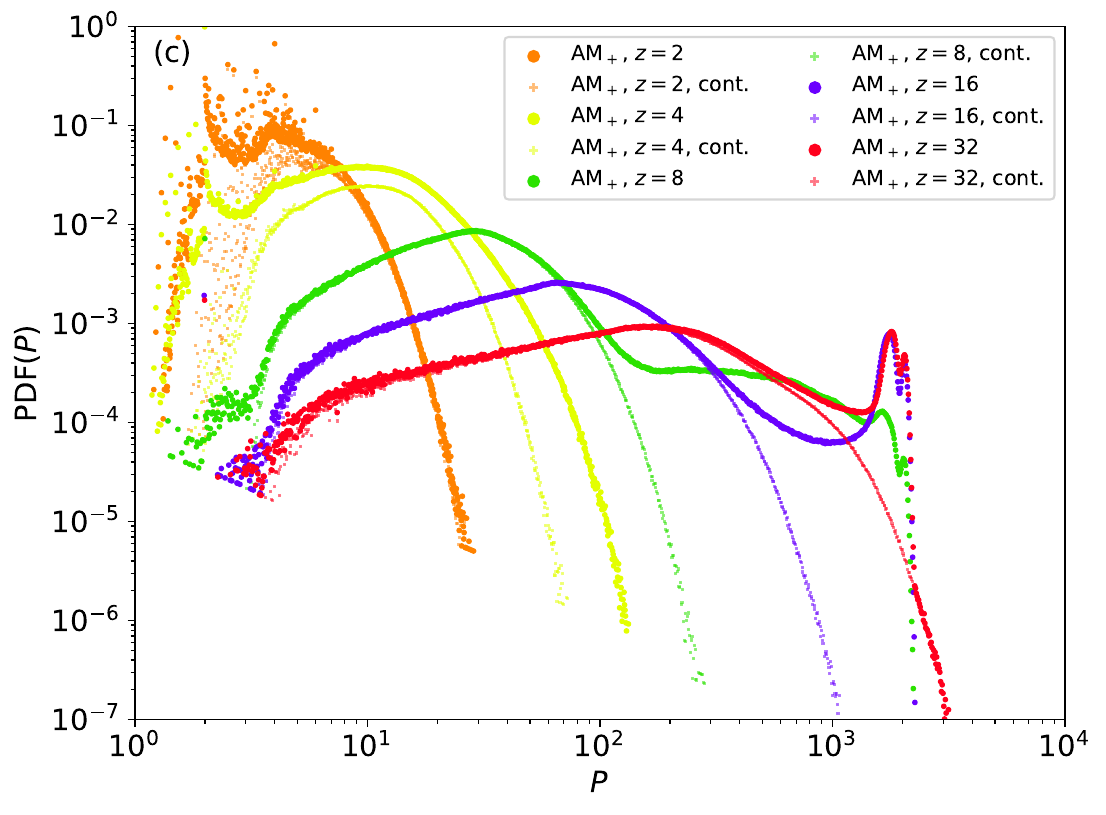} 
            \label{fig:subfig3}
        \end{subfigure}%
        \begin{subfigure}{0.5\textwidth}
            \centering
            \includegraphics[width=\linewidth]{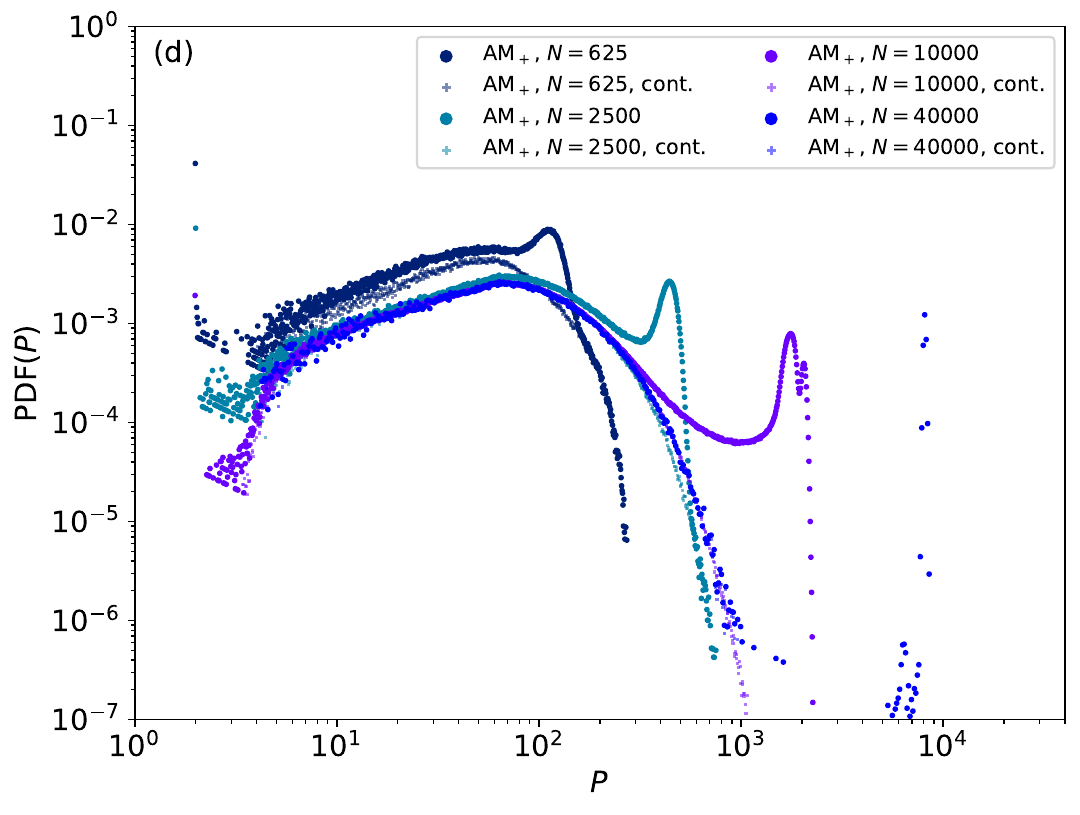}
            \label{fig:subfig4}
        \end{subfigure}
    \end{minipage}

    \begin{minipage}{\textwidth}
        \centering
        \begin{subfigure}{0.5\textwidth}
            \centering
            \includegraphics[width=\linewidth]{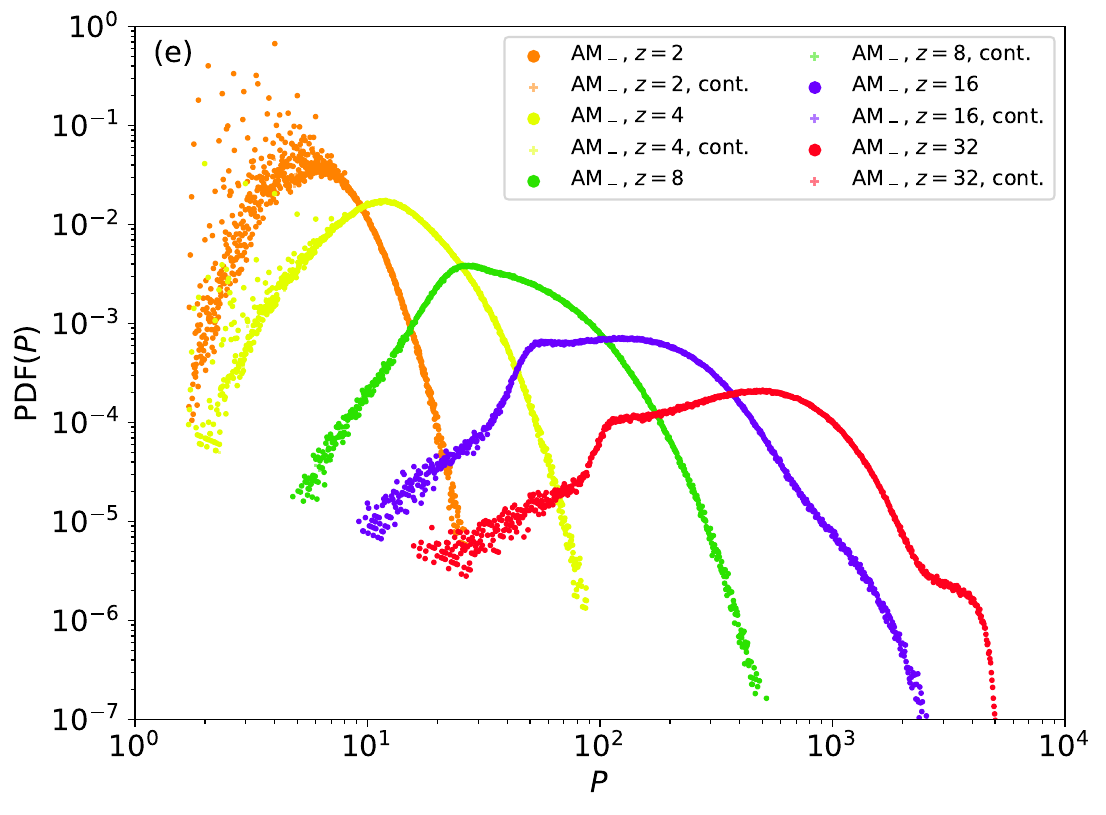} 
            \label{fig:subfig5}
        \end{subfigure}%
        \begin{subfigure}{0.5\textwidth}
            \centering
            \includegraphics[width=\linewidth]{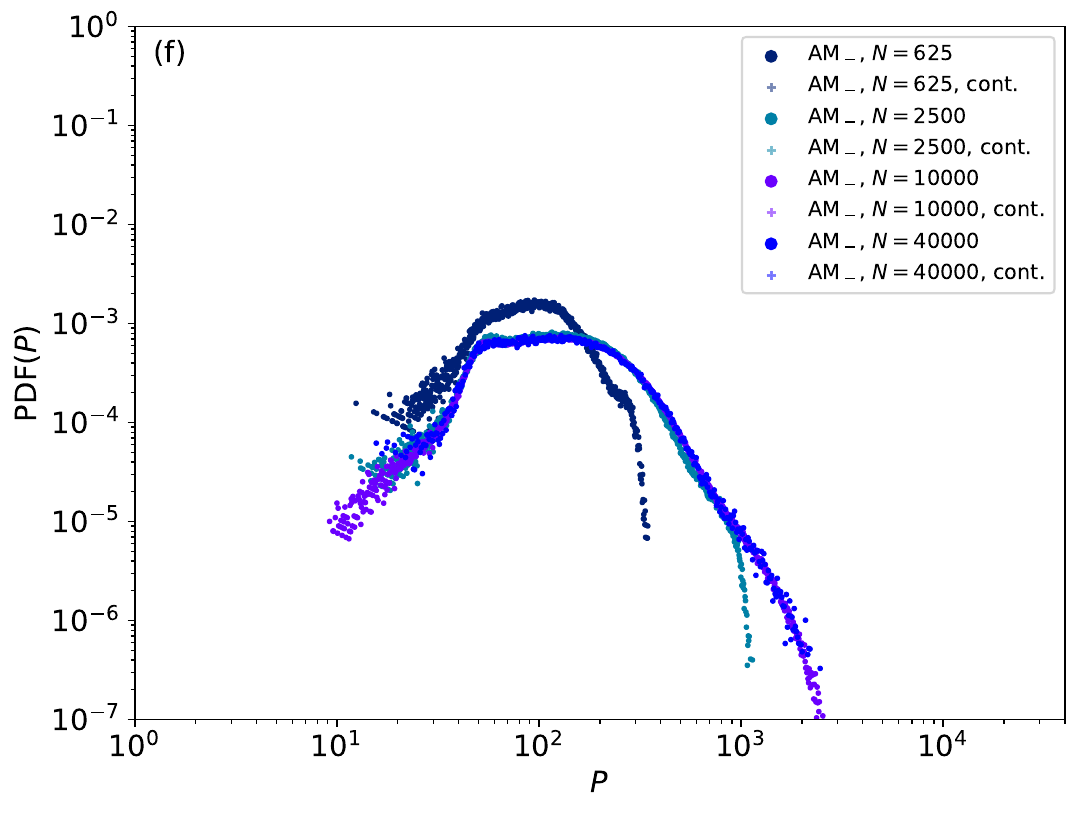} 
            \label{fig:subfig6}
        \end{subfigure}
    \end{minipage}

    \caption{Distribution of the participation ratio for fixed $N=10^4$ (left) and fixed $z=16$ (right). The first row shows the LM, the second row the AM with positive eigenvalues, and the third row the AM with negative eigenvalues. The data points with "+" markers are obtained when only modes with noninteger energies are taken into account, and they are only visible when they deviate considerably from the full curve, where they give rise to curve segments that appear lighter than the rest and lie below the main curve. They are not visible at all in Figs.(e) and (f). The gray dashed line in (a) is the size distribution of graph components for $z=2$. The continuous part of the spectrum for $z=2$ coincides with the lower branch of the curve.}
    \label{fig:PPDF_scaled}
\end{figure}

For the AM, the data show a simpler structure, with the continuous part fully localized and finite-size effects occurring only if the system size is in or below the cutoff region of the probability distribution of the participation ratio (see Fig.~\ref{fig:PPDF_scaled}(d) and (f)). The discrete part of the data occurs only for positive energies, as all orbits have the energy value 1 (or 0). Large components can contain several of these orbits, and obviously the algorithm that determines the eigenvectors yields linear combinations of the minimal modes that are fully localized, see Fig.~\ref{fig:EV_AM}(d). Compared to the LM (Fig.~\ref{fig:EV_LM}(d)), the participation ratios of these modes are larger, since the extent of degeneracy is greater. So the right-hand end of the distribution for the AM is due to degenerate orbits, i.e., the discrete part of the spectrum, while the right-hand end of the distribution for the LM is due to the system-spanning modes and widely extended modes of the continuous part of the spectrum, with the discrete part causing the "noses" in the central part of the curves (compare in Fig.~\ref{fig:PPDF_scaled}(a) and (c) the thinner curves, where the discrete part of the spectrum is omitted, with full curves.)

\subsection{Contribution of different component sizes to the PPDF}
\begin{figure}[H]
    \centering
    \begin{minipage}{\textwidth}
        \centering
        \begin{subfigure}{0.5\textwidth}
            \centering
            \includegraphics[width=\linewidth]{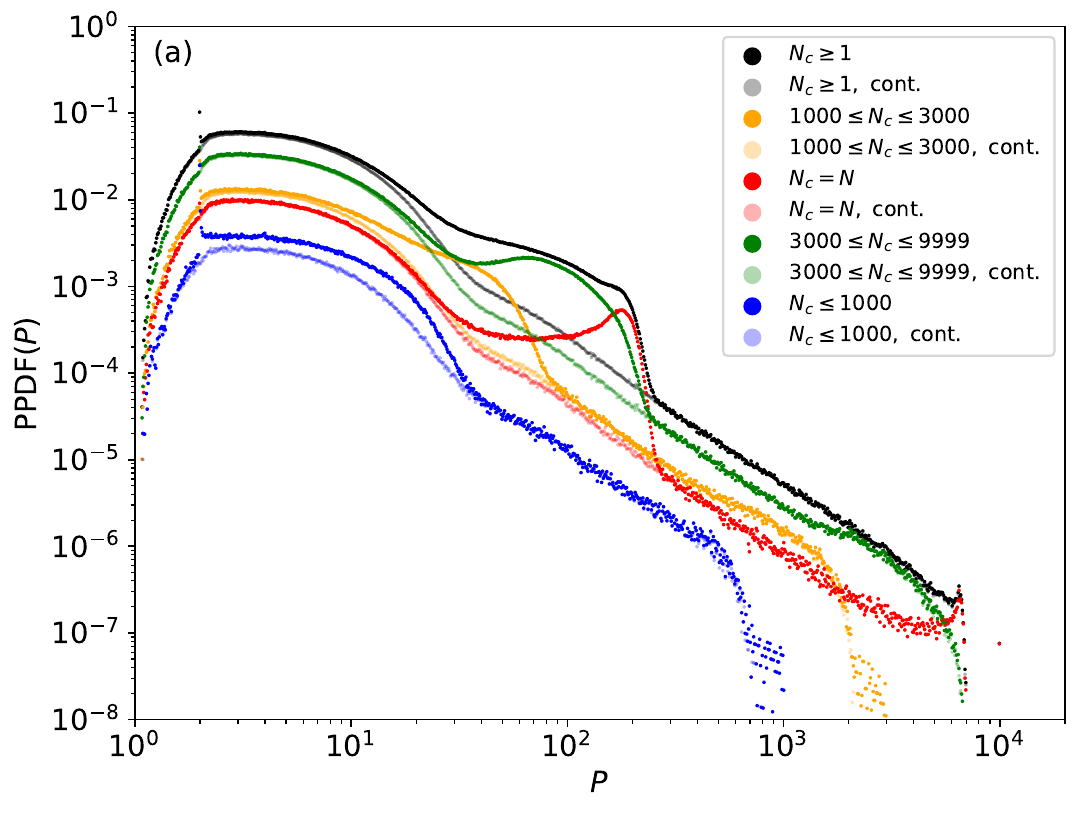} 
            \label{fig:subfig1}
        \end{subfigure}%
        \hfill
        \begin{subfigure}{0.5\textwidth}
            \centering
            \includegraphics[width=\linewidth]{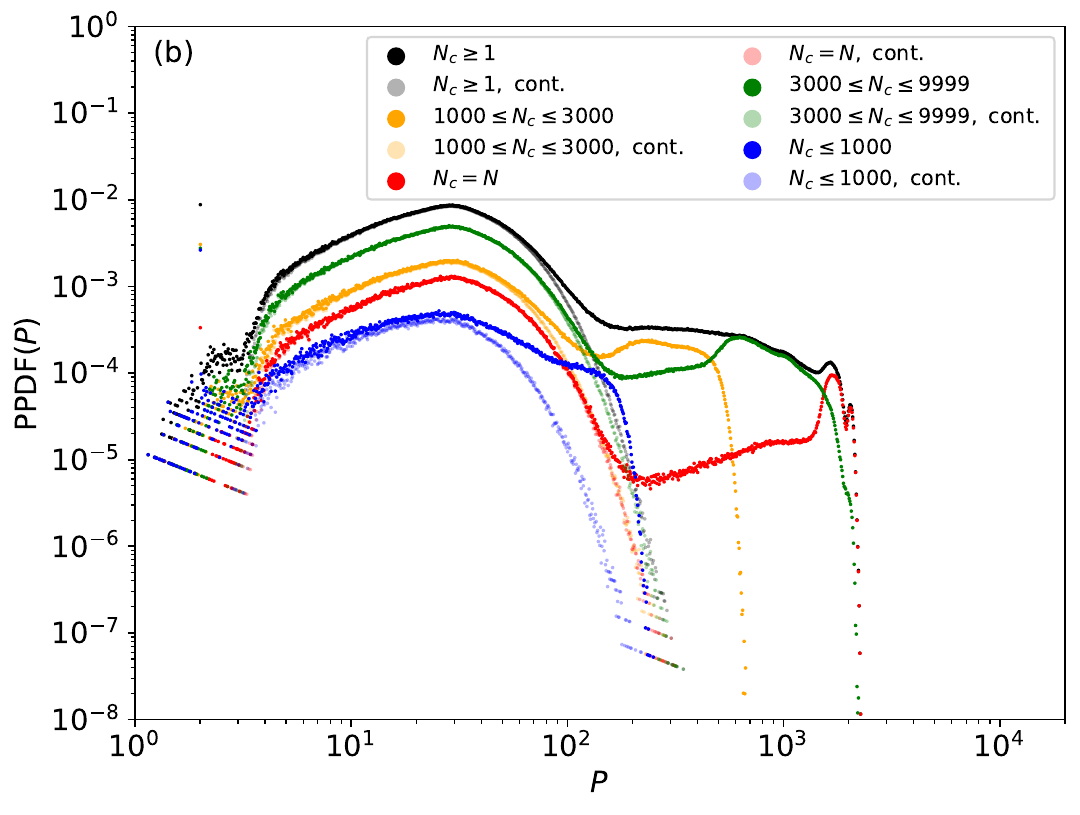} 
            \label{fig:subfig2}
        \end{subfigure}
    \end{minipage}

    \caption{Contribution of different component sizes to the probability distribution of the participation ratio (PPDF) for the LM (a) and the AM for positive energies (b), for $N=10^4$ and $z=8$. The lighter colors indicate again the modes with noninteger energies.}
    \label{fig:PPDF_components}
\end{figure}

Figure \ref{fig:PPDF_components} shows the contributions of components of different sizes to the PPDF for $z=8$, where the ensemble of systems has system-spanning components as well as smaller components. The lighter colors show only the modes with noninteger energy values. For the LM, we see  that only the system-spanning component (red) contributes to the peak  at the right-hand end of the distribution, and that the hump due to orbits with integer-valued energies is composed of a "nose" from the system-spanning component and a broader peak for large but not system-spanning components (green). For the smaller components, the contribution of modes with integer-valued energies is mostly visible for smaller $P$, where the light blue curve is below the intense blue curve. At $P=2$, there is a distinct peak due to orbits that are located on two nodes. 
For the AM, we see similar effects, but the participation ratios of the modes with integer energies are much larger than for the LM, due to the higher degree of degeneracy of the orbits, as explained above. 

\subsection{Dependence of the participation ratio on  the energy}

\begin{figure}[H]
    \centering
    \begin{minipage}{\textwidth}
        \centering
        \begin{subfigure}{0.5\textwidth}
            \centering
            \includegraphics[width=\linewidth]{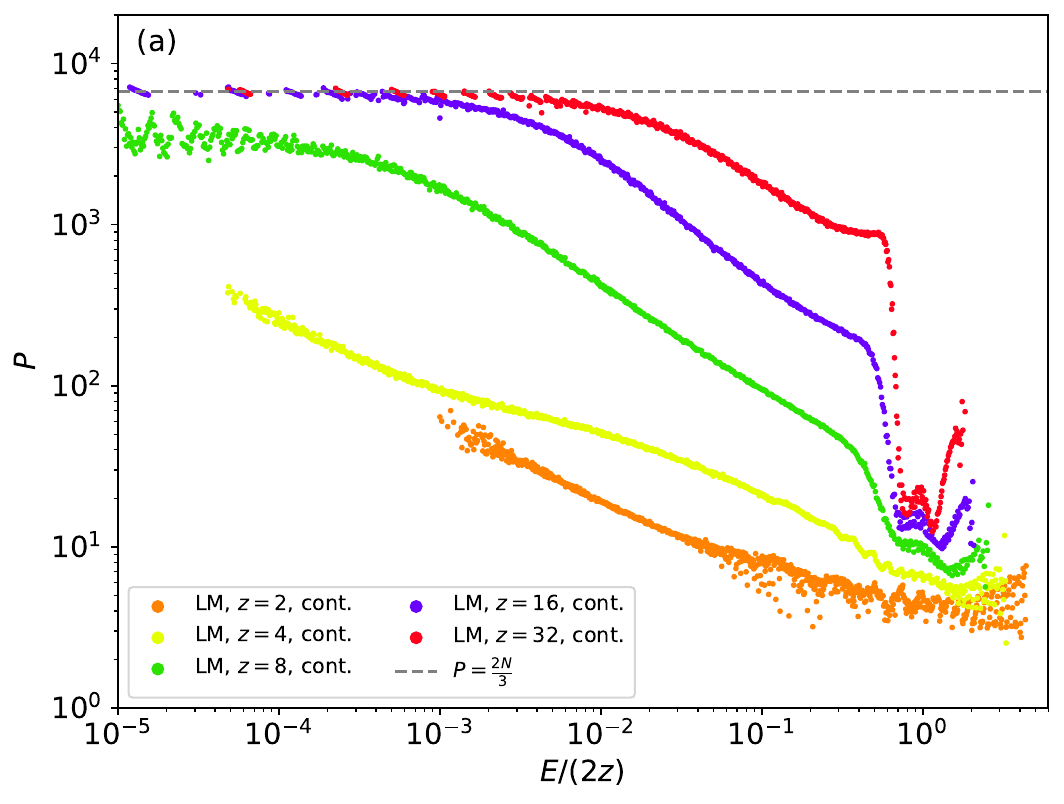} 
            \label{fig:subfig1}
        \end{subfigure}%
        \begin{subfigure}{0.5\textwidth}
            \centering
            \includegraphics[width=\linewidth]{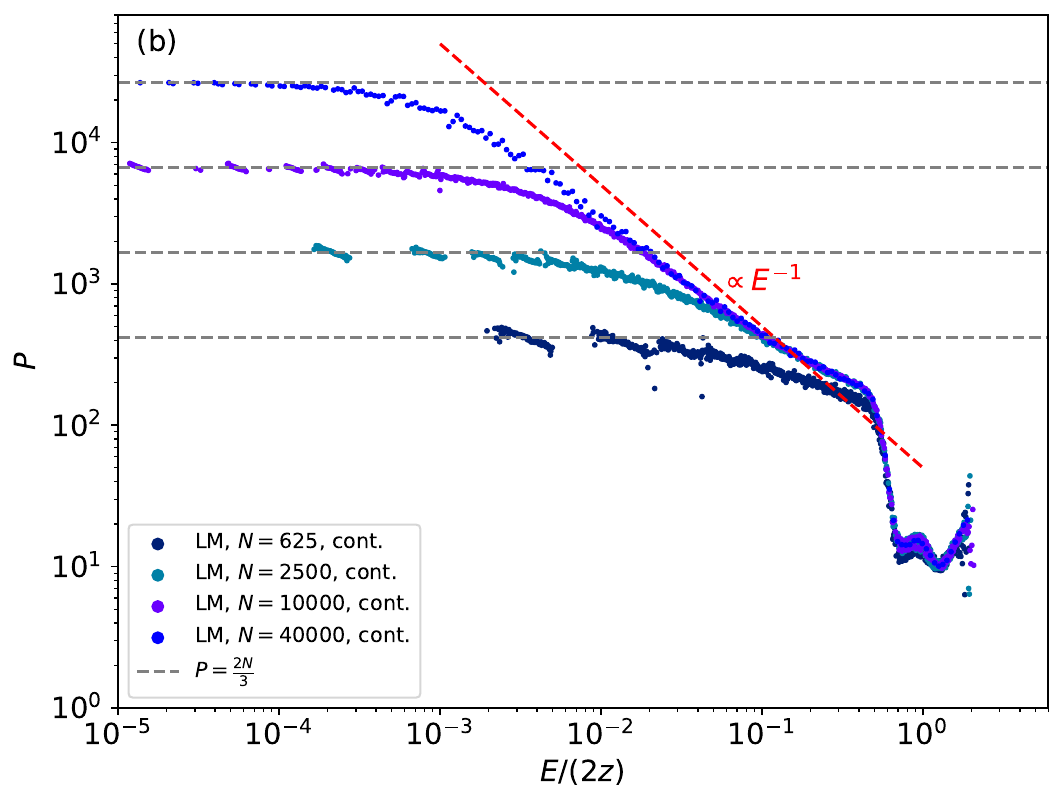} 
            \label{fig:subfig2}
        \end{subfigure}
    \end{minipage}
    \begin{minipage}{\textwidth}
        \centering
        \begin{subfigure}{0.5\textwidth}
            \centering
            \includegraphics[width=\linewidth]{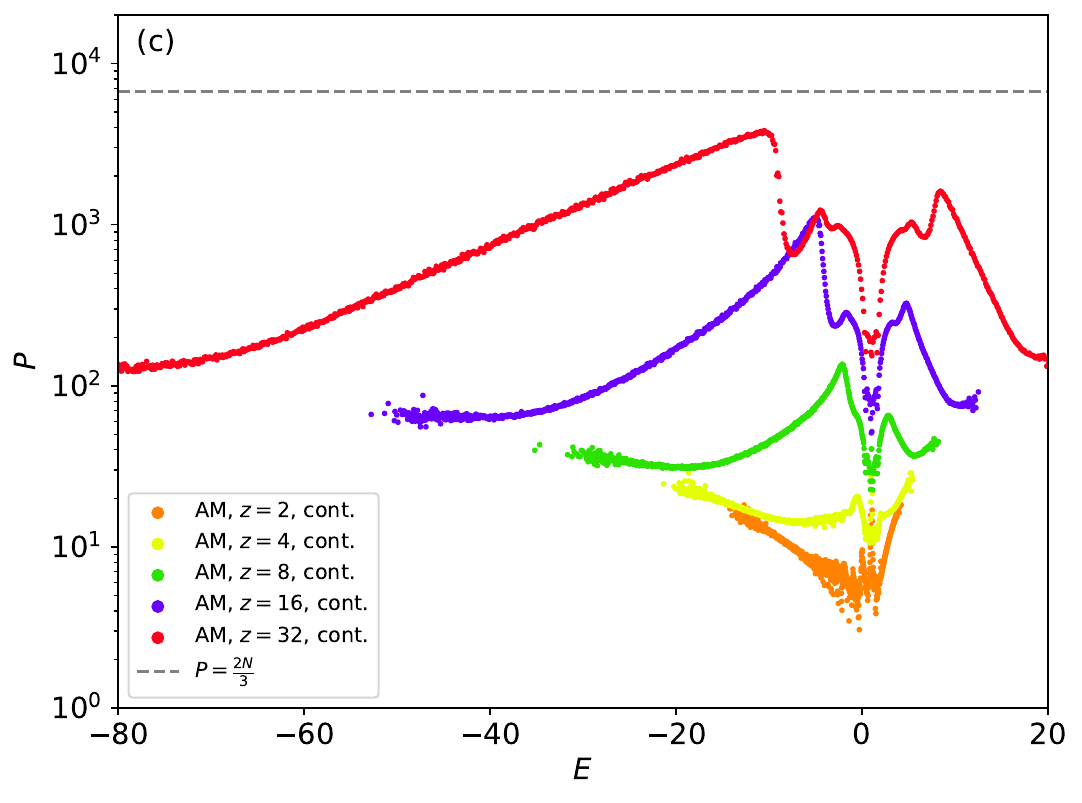} 
            \label{fig:subfig3}
        \end{subfigure}%
        \begin{subfigure}{0.5\textwidth}
            \centering
            \includegraphics[width=\linewidth]{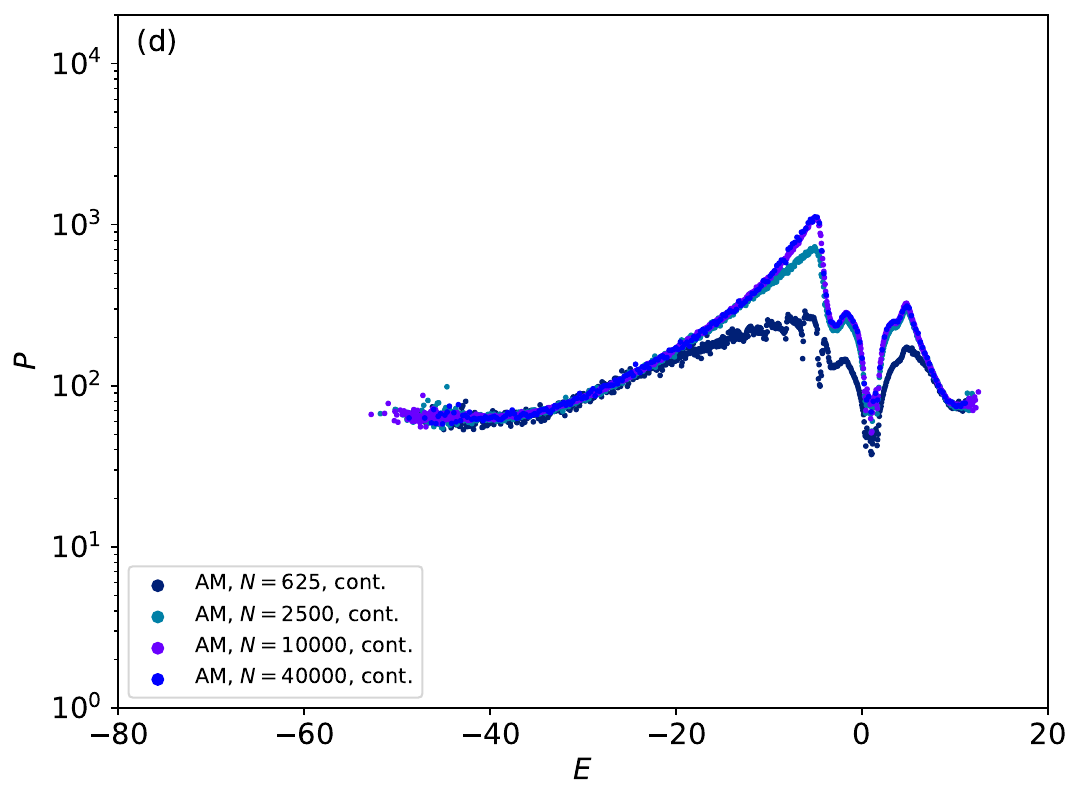}
            \label{fig:subfig4}
        \end{subfigure}
    \end{minipage}
    \caption{Dependence of the participation ratio on the energy for (a) the LM with $N=10^4$, (b) the LM with $z=16$, (c) the AM with $N=10^4$, and (d) the AM with $z=16$. The grey dashed line indicates the participation ratio of sine.}
    \label{fig:EvP}
\end{figure}

For the LM, the participation ratio increases with decreasing $E$, as extended modes are those with long wavelengths and, therefore, small energies. The slope becomes steeper as the system size becomes larger, see Fig.~\ref{fig:EvP}(b) and we expect the data to eventually approach a power law with slope -1. This is the slope that we observed in the 1D lattice with random couplings \cite{schaeferScalingBehaviourLocalised2024}.   For even smaller $E$, the $P$ values approach the value $2N/3$, which applies to a sine wave. This value is not seen for the smaller $z$ values 2 and 4, since the cutoff in component size does not allow modes with sufficiently large extension and wavelength; see Fig.~\ref{fig:EvP}(a). For energy values around $2z$, the data show a peak which is due to the orbits, and which vanishes when only the continuous part of the spectrum (i.e. noninteger values of $E$) is considered. This is also the region where the density of states has its maximum. 

For the AM, we see also first an increase of $P$ with decreasing $|E|$, at least for the larger $z$ values, but at small $|E|$ there is a funnel centered around $E=1$, which is the energy value of the orbits. Near $E=1$, there are well localized modes with small participation ratios. This non-monotonous behavior of $P$ as function of $E$ leads to several eigenvectors with vastly different energies having the same participation ratio, as shown above in Fig.~\ref{fig:EV_AM_same_P}.

\subsection{Degeneracy of eigenvectors for small $z$}\label{sec:blockdiagonal}
\begin{figure}[H]
    \centering
    \begin{minipage}{\textwidth}
        \centering
        \begin{subfigure}{0.5\textwidth}
            \centering
            \includegraphics[width=\linewidth]{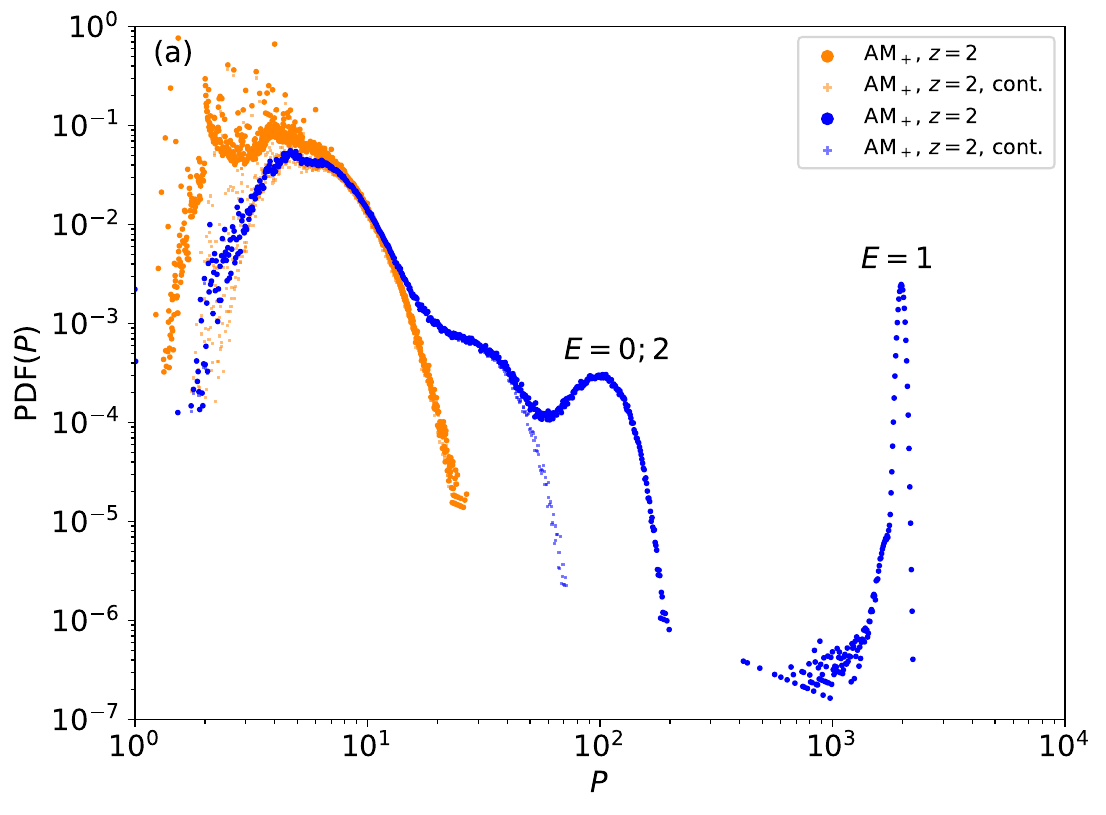} 
            \label{fig:subfig1}
        \end{subfigure}%
        \begin{subfigure}{0.5\textwidth}
            \centering
            \includegraphics[width=\linewidth]{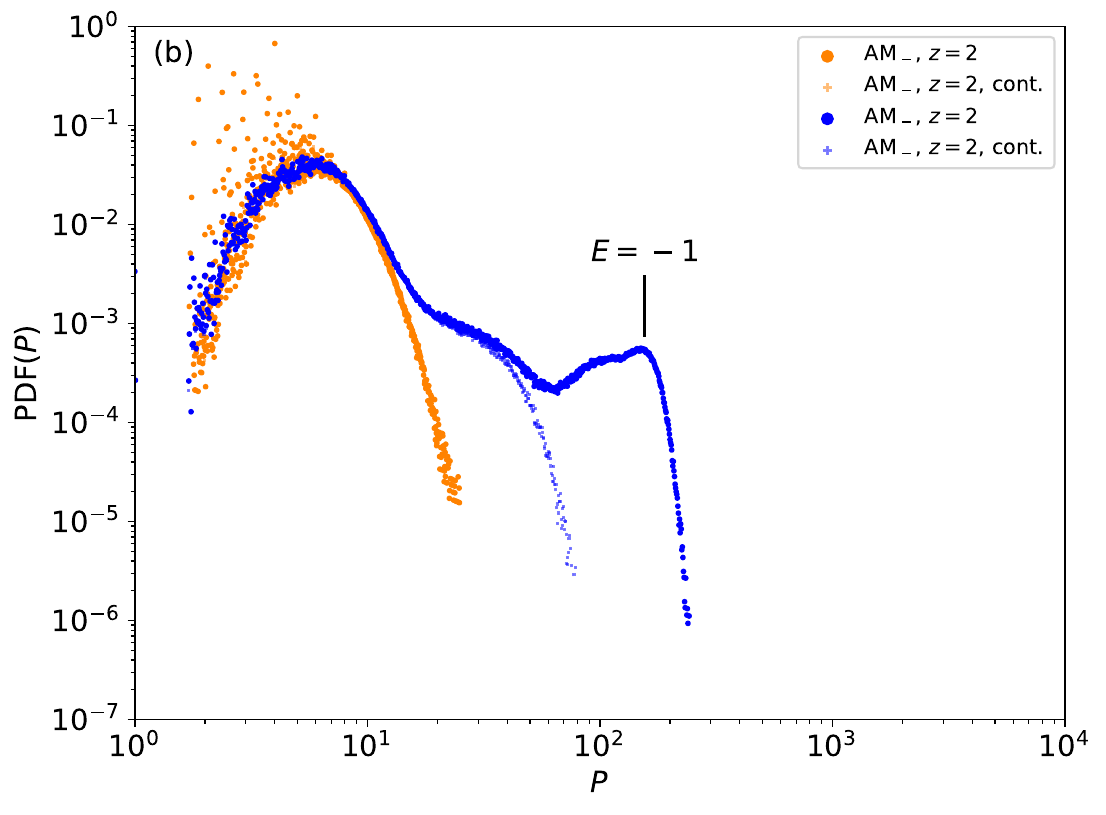} 
            \label{fig:subfig2}
        \end{subfigure}
    \end{minipage}
    \begin{minipage}{\textwidth}
        \begin{subfigure}{0.5\textwidth}
            \centering
            \includegraphics[width=\linewidth]{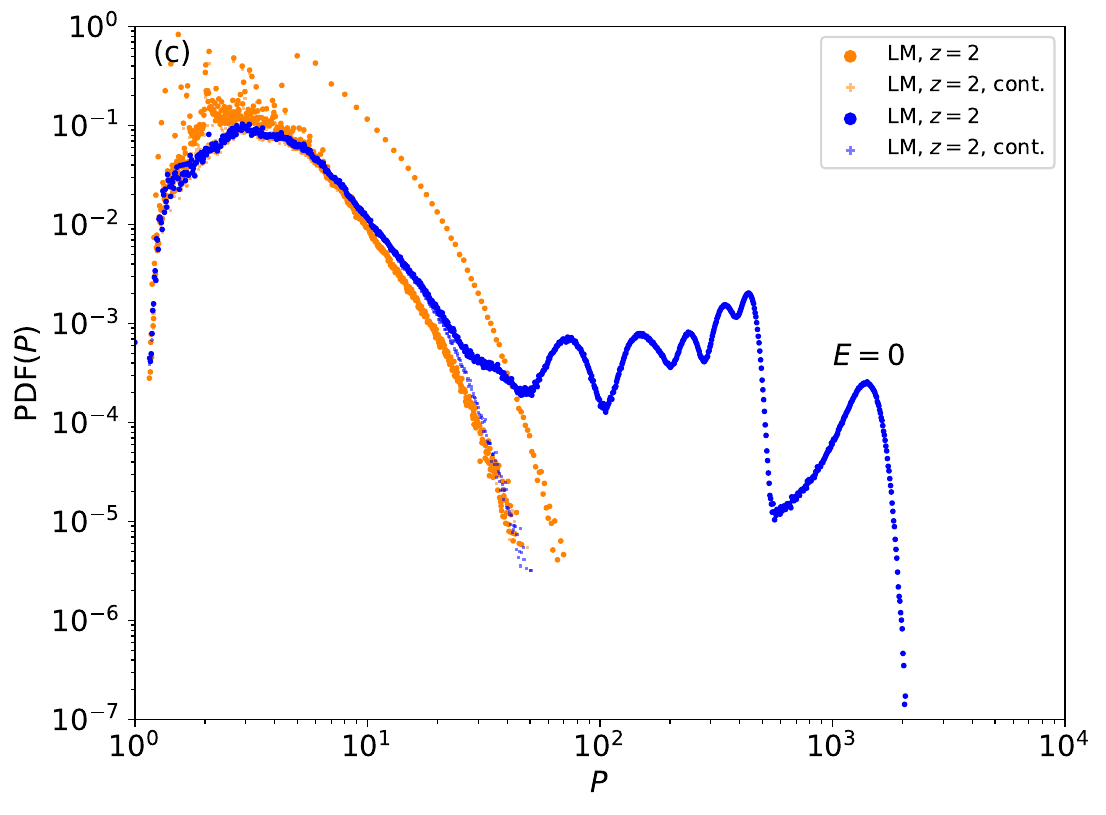}
            \label{fig:subfig3}
        \end{subfigure}%
    \end{minipage}

    \caption{Distribution of the participation ratio for fixed $N=10^4$ and $z=2$. The blue data points are obtained from systems in which the node positions have not been ordered before diagonalization whereas the orange ones have been ordered. (a) shows the PPDF of the AM for positive, (b) for negative eigenvalues, and (c) the PPDF for the LM.}
    \label{fig:PPDF_disordered}
\end{figure}

For $z=2$ and $z=4$, the system is made up of many small components. This leads to a high degree of degeneracy of eigenvectors, not only due to orbits, but also due to identical components. The diagonalization algorithm that produced all previous figures was based on an ordering of nodes such that the Hamiltonian had block diagonal form and all eigenvectors were confined to one component. When we do not sort the nodes before determining the eigenvectors, the algorithm generates linear combinations of degenerate eigenvectors from different components. Fig.~\ref{fig:PPDF_disordered} shows how the probability distribution of the participation ratio changes when the nodes are not sorted. 
The participation ratios associated with the continuous part of the spectrum now show a hump for the AM, for both signs of the energy. This must be due to identical components. For the LM, there is a slight shoulder, but the effect is much less pronounced than for the AM.  The most striking changes occur for the orbits. The degeneracy of the eigenvalues associated with orbits leads for the AM for positive energies to a pronounced peak at large $P$, which corresponds to the eigenvalue $E=1$, and to a second peak at somewhat smaller $P$, which corresponds to the eigenvalue 0 and occurs in the graphs for positive and negative energies, depending on the sign of the floating-point number that the algorithm yields as eigenvalue. For the LM, there is a series of peaks that are due to the orbits, with the rightmost peak being associated with the eigenvalue 0, and with  the eigenvalues increasing from right to left (but not matching exactly the different peaks).

\section{Discussion and Conclusion}

Our comprehensive evaluation of the localization properties of the eigenmodes of the Laplace and adjacency matrix of one-dimensional random geometric graphs shows how component size distribution, mean degree, network motifs (orbits), and system size affect the distribution of the participation ratio and the extent of localization. Since the model based on the Laplace matrix (LM) can be interpreted as a dynamical model with a conserved quantity (see Eq.~\eqref{LM_timedependent}), this model has system-spanning eigenmodes that have the shape of sine waves (see Fig.~\ref{fig:EV_LM}(a)) and give rise to the power law with exponent $-0.5$ in the density of states (Fig.~\ref{fig:DOS}(b)), just as in the ordered model (Fig.~\ref{fig:DOSDGG}(b)). In the probability distribution of the participation ratio, these modes lead to the final increase at the right-hand end of the curves (Fig.~\ref{fig:PPDF_scaled}(a) and (b)), with the corresponding $P$ values being 2/3 of the system size (see Fig.~\ref{fig:EvP}(a) and (b)). In contrast, the model that is based on the adjacency matrix (AM) has a cutoff in the distribution of the participation ratio that becomes independent of the system size when the system size is large enough and the discrete part of the spectrum is ignored (see Fig.~\ref{fig:PPDF_scaled}(c)-(f)). 

The discrete part of the spectrum is mostly due to orbits, i.e., eigenmodes that are fully localized on small network motifs with integer eigenvalues. Due to the high occurrence of these orbits and the degeneracy of their eigenvalues, they give rise to large values of the participation ratio, in particular for the AM, where all orbits have the eigenvalues 0 or 1 (see Figs.~\ref{fig:PPDF_disordered} and \ref{fig:PPDF_scaled}). Pictures of such degenerate eigenmodes are shown in Figs.~\ref{fig:EV_LM}(c) and \ref{fig:EV_AM}(c). For the LM, the peak of the probability distribution of the participation ratios due to orbits moves to smaller $P$ values with increasing mean degree $2z$ for the LM as the eigenvalues associated with orbits become larger and their distribution broader (see Fig.~\ref{fig:PPDF_scaled}(a)), whereas for the AM the position of the peak due to orbits is almost independent of $z$ (see Fig.~\ref{fig:PPDF_scaled}(c)), indicating that the proportion of orbit eigenmodes among all eigenmodes remain of the same order as $z$ is changed. This agrees with the finding of Nyberg et al.~\cite{nybergLaplacianSpectraRandom2014} that the proportion of eigenvalues due to type-I orbits is of the order of 1/3 if $z\gg1$. 

For $z=2$ and 4, the size distribution of network components is such that the graph is composed of many components, and the cutoff in the component size distribution is much smaller than the system size and determines also the cutoff in the probability distribution of the participation ratio (see Figs.~\ref{fig:PPDF_scaled}(a) and (c)). The case $z=8$ is a limit case, where part of the graph realizations consist of a single, system-spanning component, and the probability distribution of the participation ratio has qualitatively different contributions from components of different sizes, see Fig.~\ref{fig:PPDF_components}. For $z=16$ and $32$, the vast majority of graph realizations consists of one component. For these values of $z$, the finite-size effects are strongest, since the distribution of the participation ratio becomes shifted to larger $P$ values (Fig.~\ref{fig:PPDF_scaled}). We explain  this by the fact that larger values of $z$ lead to larger sets of nearby nodes that are fully connected to each other. With respect to the disorder in the system, such fully connected groups of nodes act as a single effective node, and the effective system size becomes smaller (when measured in units of these effective nodes). This effect is most clearly visible in Fig.~\ref{fig:PPDF_scaled}(a) and (b), where the right-hand hump hides the power law expected for larger $N$, and this effect is stronger for larger $z$. The increase of finite-size effects with increasing $z$ is also visible in Fig.~\ref{fig:DOS}(b), where the $z=8$ data agree with the slope $-0.5$ over a much larger interval than the data for $z=16$ or $z=32$. The finite-size effects for $z=16$ and $N=10^4$ lead also to the system-spanning mode shown for the AM (Fig.~\ref{fig:EV_AM}(b)). 

With increasing $z$, the density of states of the LM splits more and more into two separate parts, i.e., the peak around $E=2z$, and the power law for small energies (see Fig.~\ref{fig:DOS}(b)). The depth of the valley between these two parts scales as $1/z^{1.5}$ for large $z$, which follows from the scaling shown in Fig.~\ref{fig:DOS}(b), which in turn follows from Eq.~\eqref{eq:DOS_correction}. This valley of the density of states leads to the valley in the probability distribution of the participation ratio to the left of the pronounced hump of the LM (Fig.~\ref{fig:PPDF_scaled}(a)). 

The local connectivity patterns of RGGs manifest themselves mainly on short distances, where the presence of orbits and strongly connected sets of nodes become visible. On larger scales, the LM and the AM become more similar to one-dimensional tight-binding models with disorder \cite{schaeferScalingBehaviourLocalised2024}, where the presence or absence of a conservation law (LM versus AM in this paper, or DM and RCM in the tight-binding models studied in \cite{schaeferScalingBehaviourLocalised2024}) determines the large-scale features of eigenmodes and their localization. Both the AM and the RCM show a unimodal distribution of participation ratios with a cutoff that becomes independent of the system size when the system size is large enough (compare the green curve in Fig.~3(a) of \cite{schaeferScalingBehaviourLocalised2024} with Fig.~\ref{fig:PPDF_scaled}(c)-(f) above and omit the part due to the orbits). The main difference between the two models is due to the fact that the RCM is bipartite, as the one-dimensional regular lattice can be decomposed into two sublattices. This leads to a symmetry of eigenmodes with respect to the sign of the energy, which is no longer present in the data above. For this reason, we show in most plots for the AM the parts for positive and negative energies separately. 

The LM and the DM both have sine-shaped system-spanning modes which resemble those of the ordered system, and their proportion increases as the square root of the system size (see, e.g., \cite{schaeferScalingBehaviourLocalised2024})). In addition, both models have extended but localized modes that resemble sine waves with a variable amplitude, see Fig.~\ref{fig:EV_LM}(b). These modes give rise to the power law with exponent -1.5 in the distribution of the participation ratios (Fig.~\ref{fig:PPDF_scaled}(b)), and -1 in the relation between $P$ and $E$ (Fig.~\ref{fig:EvP}(b)), which are the same exponents as reported in \cite{schaeferScalingBehaviourLocalised2024} for the DM. 

To conclude, while the Laplace and adjacency matrix encode the same structural information of the RGGs, the localization properties of their eigenmodes are qualitatively different, due to the LM having a conserved quantity, while the AM has not. On short scales, the existence of orbits and strongly connected sets of nodes, and the component size distribution determine the features of the eigenmodes, while on large distances the two models become similar to their tight-binding counterpart when the mean degree is large enough that the graph consists of only one component.

\subsection*{Acknowledgements}
We acknowledge financial support for this project by DFG grant number Dr300/15.

\subsection*{Data availability}
The data that support the findings in this article will be publicly available upon publication of this article and the link will be added here.

\printbibliography

@book{vanmieghemGraphSpectraComplex2023,
	title={Graph spectra for complex networks},
  author={Van Mieghem, Piet},
  year={2023},
  publisher={Cambridge university press}
}

@thesis{nybergLaplacianSpectraRandom2014,
	title = {The Laplacian spectra of random geometric graphs},
	url = {https://uh-ir.tdl.org/bitstream/handle/10657/1646/NYBERG-DISSERTATION-2014.pdf?sequence=1&isAllowed=y},
	institution = {University of Houston},
	type = {phdthesis},
	author = {Nyberg, Amy},
	urldate = {2023-04-28},
	date = {2014-12},
}

@article{nybergMesoscopicStructures2015,
	title = {Mesoscopic structures and the Laplacian spectra of random geometric graphs},
	volume = {3},
	issn = {2051-1310},
	url = {https://doi.org/10.1093/comnet/cnv004},
	doi = {10.1093/comnet/cnv004},
	pages = {543--551},
	number = {4},
	journaltitle = {Journal of Complex Networks},
	author = {Nyberg, Amy and Gross, Thilo and Bassler, Kevin E.},
	date = {2015-03},
}

@article{schaeferScalingBehaviourLocalised2024,
	title = {The scaling behaviour of localised and extended states in one-dimensional tight-binding models with disorder},
	volume = {100},
	issn = {0031-8949, 1402-4896},
	url = {https://iopscience.iop.org/article/10.1088/1402-4896/ad9e44},
	doi = {10.1088/1402-4896/ad9e44},
	pages = {015991},
	number = {1},
	journaltitle = {Physica Scripta},
	shortjournal = {Phys. Scr.},
	author = {Schaefer, Luca and Drossel, Barbara},
	urldate = {2025-02-28},
	date = {2025-01-01},
	langid = {english},
}

@article{barthelemySpatialNetworks2011a,
	title = {Spatial Networks},
	volume = {499},
	issn = {03701573},
	url = {http://arxiv.org/abs/1010.0302},
	doi = {10.1016/j.physrep.2010.11.002},
	pages = {1--101},
	number = {1},
	journaltitle = {Physics Reports},
	shortjournal = {Physics Reports},
	author = {Barthelemy, Marc},
	urldate = {2024-10-31},
	date = {2011-02},
	langid = {english},
	eprinttype = {arxiv},
	eprint = {1010.0302 [cond-mat]},
}

@article{andersonAbsenceDiffusionCertain1958,
  title = {Absence of {{Diffusion}} in {{Certain Random Lattices}}},
  author = {Anderson, P. W.},
  date = {1958-03-01},
  journaltitle = {Physical Review},
  shortjournal = {Phys. Rev.},
  volume = {109},
  number = {5},
  pages = {1492--1505},
  issn = {0031-899X},
  doi = {10.1103/PhysRev.109.1492},
  url = {https://link.aps.org/doi/10.1103/PhysRev.109.1492},
  urldate = {2023-02-15},
  langid = {english}
}

@article{hirotsugamatsudaLocalizationNormalModes1970,
	title = {Localization of Normal Modes and Energy Transport in the Disordered Harmonic Chain},
	url = {https://watermark.silverchair.com/45-56.pdf?token=AQECAHi208BE49Ooan9kkhW_Ercy7Dm3ZL_9Cf3qfKAc485ysgAAA4QwggOABgkqhkiG9w0BBwagggNxMIIDbQIBADCCA2YGCSqGSIb3DQEHATAeBglghkgBZQMEAS4wEQQMBzCt8BQF5TS5AzmlAgEQgIIDN8GrdPoZueudNhi47XsrjV1ROnHhmy8w-6UCzGQDWJ6Hqyvn7uCxzLw5dY3djHk4uH9joeb2MgNhoxxjIbG6AZ2lxUO7Zz-6qlBOhQuDTtpMpv8sONiulYIrxYJJ5wPuB-6l2KY4A-Ceh6pQI8cenGw6-ge289j5AiwpDy34jSrXzy1ikXXc56wsx1drNEr7uGqSemRbXOLmkpXyfSJ_xz6RcQyljGHoIatt6SorraiC9fX5oslFE8ttTsWqXlfLv5jmb42hUQzZJwP62OE2w9mm-hpkM2b-7UdSCcj6YLHR2ZI20UTTGRYHHr3zvCBrWhDhzWAaLj0MaS3mEv4GnppFCFFGXqDwhFKrnq5u2qPJdrv6mdwPBZdeqvq9xusye6AJS7fOPEQqbU9frWFhslBULuVVKb9XBcqqV5dT-1MioMxukmWCD1GwBkBUGzcd676JVSugtJ4qLaPWEzOVnkrae1pLGwR9ExhJ4dntjOBYjiiT8qOG4rHgfrTwTtjkwOIP3h7KeSHYNIUuFBZue0_Scw57vMIhfKMhlkOTRcY289xUMu0avNUuD8OlNZVB5IA3JK18L1hovZVtGE5qFDjogrihnprAd731ATXIqWNd1e5ysfL2vwavZLcQ95ZW27N9RQHt7NM6j_RqBLOiR_c-PK51tb5DjXIr9WnJGjal_3byZ7TNdrr-j555Bdinl611b39mEevIMmRVL45cfNf8Cy0pQqXa7C9fQgMw9sFDPJYyqNlZ4lD4R8lp4rK05O4G4Bo2ubM_C71vZyMJIr1ektZmfaToZdXdG5CCM1kUDAsA_VVh8ZCRqjKT4tJ6rO7CooPvGx-ydpTR7Iwbr_DCMoZfMYhynIgDuoTXWUwWsaGwoRuWvxKU-ismwVnQTfhO3BLuR_bjLEjPoxCGXmtVVWHEtLkYo-Qw5ZbOv3QI7J9c5IBMXYOkxHqhsX8XiDSkatsPNueZL5pj_BfsxAZyE4lJlj2a9_359yk3Oibzic3dnKm7lfE-wAXwEDQuqSKMJqAFWZcEg_cEewAxdZp8Ry7-qhDHv28VN1ByibT1P8tlj9bJDwrZgX4rg-kUm7jawxNivFI},
	number = {45},
	journaltitle = {Supplement of the Progress of Theoretical Physics},
	author = {{Hirotsuga Matsuda} and {Kazushige Ishii}},
	urldate = {2023-09-22},
	date = {1970},
}

@article{dominguez-adameDelocalizedVibrationsClassical1993,
	title = {Delocalized vibrations in classical random chains},
	volume = {48},
	issn = {0163-1829, 1095-3795},
	url = {https://link.aps.org/doi/10.1103/PhysRevB.48.6054},
	doi = {10.1103/PhysRevB.48.6054},
	pages = {6054--6057},
	number = {9},
	journaltitle = {Physical Review B},
	shortjournal = {Phys. Rev. B},
	author = {Dom\'{i}nguez-Adame, Francisco and Maci\'{a}, Enrique and S\'{a}nchez, Angel},
	urldate = {2023-12-08},
	date = {1993-09-01},
	langid = {english},
}

@article{delyonOnedimensionalWaveEquations1983,
	title = {One-dimensional wave equations in disordered media},
	volume = {16},
	issn = {0305-4470, 1361-6447},
	url = {https://iopscience.iop.org/article/10.1088/0305-4470/16/1/012},
	doi = {10.1088/0305-4470/16/1/012},
	pages = {25--42},
	number = {1},
	journaltitle = {Journal of Physics A: Mathematical and General},
	shortjournal = {J. Phys. A: Math. Gen.},
	author = {Delyon, F and Kunz, H and Souillard, B},
	urldate = {2025-07-10},
	date = {1983-01-11},
	langid = {english},
	note = {Publisher: {IOP} Publishing},
}

@article{paytonDynamicsDistortedHarmonic1967,
	title = {Dynamics of Distorted Harmonic Lattices. {II}. The Normal Modes of Isotopically Disordered Binary Lattices},
	volume = {156},
	issn = {0031-899X},
	url = {https://link.aps.org/doi/10.1103/PhysRev.156.1032},
	doi = {10.1103/PhysRev.156.1032},
	pages = {1032--1038},
	number = {3},
	journaltitle = {Physical Review},
	shortjournal = {Phys. Rev.},
	author = {Payton, Daniel N. and Visscher, William M.},
	urldate = {2024-03-13},
	date = {1967-04-15},
	langid = {english},
}

@article{p.deanVibrationsTwodimensionalDisordered,
	title = {Vibrations of two-dimensional disordered lattices},
	author = {{P. Dean} and {M.D. Bacon}},
	langid = {english},
    journaltitle = {Proc. R. Soc. Lond. A},
    year = {1965},
    number = {283},
    pages = {64-82},
}

@article{mcgrawLaplacianSpectraDiagnostic2008,
  title = {Laplacian Spectra as a Diagnostic Tool for Network Structure and Dynamics},
  author = {McGraw, Patrick N. and Menzinger, Michael},
  date = {2008-03-04},
  journaltitle = {Physical Review E},
  shortjournal = {Phys. Rev. E},
  volume = {77},
  number = {3},
  pages = {031102},
  issn = {1539-3755, 1550-2376},
  doi = {10.1103/PhysRevE.77.031102},
  url = {https://link.aps.org/doi/10.1103/PhysRevE.77.031102},
  urldate = {2023-02-15},
  langid = {english}
}

@article{zhuLocalizationsComplexNetworks2008,
	title = {Localizations on complex networks},
	volume = {77},
	rights = {http://link.aps.org/licenses/aps-default-license},
	issn = {1539-3755, 1550-2376},
	url = {https://link.aps.org/doi/10.1103/PhysRevE.77.066113},
	doi = {10.1103/physreve.77.066113},
	number = {6},
	journaltitle = {Physical Review E},
	shortjournal = {Phys. Rev. E},
	author = {Zhu, Guimei and Yang, Huijie and Yin, Chuanyang and Li, Baowen},
	urldate = {2025-07-10},
	date = {2008-06-23},
	langid = {english},
	note = {Publisher: American Physical Society ({APS})},
}

@book{penroseRandomGeometricGraphs2003,
	location = {Oxford ; New York},
	title = {Random geometric graphs},
	isbn = {978-0-19-850626-3},
	series = {Oxford studies in probability},
	number = {5},
	publisher = {Oxford University Press},
	author = {Penrose, Mathew},
	date = {2003},
	langid = {english},
	note = {{OCLC}: ocm52603896},
}

@article{spielmanSpectralAlgebraicGraph,
  title={Spectral graph theory},
  author={Spielman, Daniel},
  journal={Combinatorial scientific computing},
  volume={18},
  number={18},
  year={2012},
  publisher={CRC Press Boca Raton, Florida}
}

@book{blackwellSpectraAdjacencyMatrices,
	title = {Spectra of adjacency matrices of random geometric graphs},
	author = {Blackwell, Paul and Edmondson-Jones, Mark and Jordan, Jonathan},
	langid = {english},
    year = {2007},
    publisher = {University of Sheffield. Department of Probability and Statistics},
}

@article{brechtelMasterStabilityFunctions2018,
	title = {Master stability functions reveal diffusion-driven pattern formation in networks},
	volume = {97},
	issn = {2470-0045, 2470-0053},
	url = {http://arxiv.org/abs/1610.07635},
	doi = {10.1103/PhysRevE.97.032307},
	pages = {032307},
	number = {3},
	journaltitle = {Physical Review E},
	shortjournal = {Phys. Rev. E},
	author = {Brechtel, Andreas and Gramlich, Philipp and Ritterskamp, Daniel and Drossel, Barbara and Gross, Thilo},
	urldate = {2023-02-15},
	date = {2018-03-19},
	langid = {english},
	eprinttype = {arxiv},
	eprint = {1610.07635 [cond-mat, physics:nlin, physics:physics]},
}

@article{hamidoucheNormalizedLaplacianSpectra2023,
	title = {On the Normalized Laplacian Spectra of Random Geometric Graphs},
	volume = {36},
	issn = {0894-9840, 1572-9230},
	url = {https://link.springer.com/10.1007/s10959-022-01158-0},
	doi = {10.1007/s10959-022-01158-0},
	pages = {46--77},
	number = {1},
	journaltitle = {Journal of Theoretical Probability},
	shortjournal = {J Theor Probab},
	author = {Hamidouche, Mounia and Cottatellucci, Laura and Avrachenkov, Konstantin},
	urldate = {2025-03-27},
	date = {2023-03},
	langid = {english},
}

@INPROCEEDINGS{hamidoucheSpectralAnalysisAdjacency2019a,

  author={Hamidouche, Mounia and Cottatellucci, Laura and Avrachenkov, Konstantin},

  booktitle={2019 57th Annual Allerton Conference on Communication, Control, and Computing (Allerton)}, 

  title={Spectral Analysis of the Adjacency Matrix of Random Geometric Graphs}, 

  year={2019},

  volume={},

  number={},

  pages={208-214},

  doi={10.1109/ALLERTON.2019.8919798}}

@article{diazSynchronizationInRandomGeometricGraphs2009,
author = {D\'{i}az-Guilera, Albert and G\'{o}mez-Garde\~{n}es, Jes\'{u}s and Moreno, Yamir and Nekovee, Maziar},
title = {Synchronization In Random Geometric Graphs},
journal = {International Journal of Bifurcation and Chaos},
volume = {19},
number = {02},
pages = {687-693},
year = {2009},
doi = {10.1142/S0218127409023044},
URL = {https://doi.org/10.1142/S0218127409023044},
eprint = {https://doi.org/10.1142/S0218127409023044}
}

@inproceedings{preciadoSpectralAnalysisVirus2009,
  author={Preciado, Victor M. and Jadbabaie, Ali},

  booktitle={Proceedings of the 48h IEEE Conference on Decision and Control (CDC)}, 

  title={Spectral analysis of virus spreading in random geometric networks}, 

  year={2009},

  volume={},

  number={},

  pages={4802-4807},

  doi={10.1109/CDC.2009.5400615}
}

@article{abrahamsScalingTheoryLocalization1979,
	title = {Scaling Theory of Localization: Absence of Quantum Diffusion in Two Dimensions},
	volume = {42},
	issn = {0031-9007},
	url = {https://link.aps.org/doi/10.1103/PhysRevLett.42.673},
	doi = {10.1103/PhysRevLett.42.673},
	shorttitle = {Scaling Theory of Localization},
	pages = {673--676},
	number = {10},
	journaltitle = {Physical Review Letters},
	shortjournal = {Phys. Rev. Lett.},
	author = {Abrahams, E. and Anderson, P. W. and Licciardello, D. C. and Ramakrishnan, T. V.},
	urldate = {2023-02-15},
	date = {1979-03-05},
	langid = {english},
}

@article{dettmannSymmetricMotifs2017,
	title = {Symmetric motifs in random geometric graphs},
	volume = {6},
	issn = {2051-1329},
	url = {https://doi.org/10.1093/comnet/cnx022},
	doi = {10.1093/comnet/cnx022},
	pages = {95--105},
	number = {1},
	journaltitle = {Journal of Complex Networks},
	author = {Dettmann, Carl P and Knight, Georgie},
	date = {2017-07},
	note = {tex.eprint: https://academic.oup.com/comnet/article-pdf/6/1/95/23676997/cnx022.pdf},
}

@article{dallRandomGeometricGraphs2002,
  title = {Random geometric graphs},
  author = {Dall, Jesper and Christensen, Michael},
  journal = {Phys. Rev. E},
  volume = {66},
  issue = {1},
  pages = {016121},
  numpages = {9},
  year = {2002},
  month = {Jul},
  publisher = {American Physical Society},
  doi = {10.1103/PhysRevE.66.016121},
  url = {https://link.aps.org/doi/10.1103/PhysRevE.66.016121}
}

@article{Dettmann_2017,
	title = {Spectral statistics of random geometric graphs},
	volume = {118},
	url = {https://dx.doi.org/10.1209/0295-5075/118/18003},
	doi = {10.1209/0295-5075/118/18003},
	pages = {18003},
	number = {1},
	journaltitle = {Europhysics Letters},
	author = {Dettmann, C. P. and Georgiou, O. and Knight, G.},
	date = {2017-05},
	note = {Publisher: {EDP} Sciences, {IOP} Publishing and Società Italiana di Fisica},
}
\end{document}